\documentclass[a4paper,twocolumn,11pt,accepted=2026-09-18]{quantumarticle}
\pdfoutput=1
\usepackage[utf8]{inputenc}
\usepackage[english]{babel}
\usepackage[T1]{fontenc}
\usepackage{amsmath}
\usepackage{hyperref}

\usepackage{tikz}
\usepackage{lipsum}

\usepackage{color,soul}
\usepackage{physics}% bra, ket
\usepackage{xcolor}
\usepackage{mathtools}
\usepackage{amssymb}
\usepackage{float}
\usepackage[numbers,sort&compress]{natbib}

\begin{document}

\title{Spinor Bose-Einstein condensate as an analog simulator of molecular bending vibrations}

\author{Ayaka Usui}
\affiliation{Departament de F\'{i}sica, Universitat Aut\`{o}noma de Barcelona, 08193 Bellaterra, Spain}
\email{ayaka.usui@uab.cat}
\orcid{0000-0002-2326-3917}
\author{Artur Niezgoda}
\affiliation{ICFO - Institut de Ciencies Fotoniques, The Barcelona Institute of Science and Technology, 08860 Castelldefels, Barcelona, Spain}
% \orcid{}
\author{Manuel Gessner}
\email{manuel.gessner@uv.es}
\affiliation{Departament de Física Teòrica and IFIC, Universitat de València-CSIC, C/Dr Moliner 50, 46100 Burjassot (Valencia), Spain}
% \orcid{}
\maketitle

\begin{abstract}
  We demonstrate that spinor Bose-Einstein condensates (BECs) can be operated as an analog simulator of the two-dimensional vibron model. This algebraic model describes bending vibrations of molecules and, in the case of triatomic molecules, exhibits two phases where linear and bent configurations are stabilised. Spinor BECs can be engineered to simulate states that correspond to linear or bent triatomic molecules, with the Wigner function of the BEC encoding information about the molecular configuration. We show how quantum simulations of the bending dynamics of linear molecules can be realised, and how preparing a linear configuration in the bent phase leads to a dynamical instability. In the dynamics triggered by the corresponding instability, a significant amount of entanglement is generated, and we characterise the dynamics with the squeezing parameter and the quantum Fisher information (QFI). The scaling of the non-Gaussian sensitivity, described by the difference between squeezing and QFI, grows with the system size once the spinor system crosses from the linear to the bent phase, thus serving as a dynamical witness for the quantum phase transition.
\end{abstract}

\section{Introduction}

Spinor Bose-Einstein condensates (BECs) have emerged as an exceptional platform for exploring many-body quantum physics under highly controlled conditions, offering access to relevant control parameters and enabling the precise preparation of diverse quantum states~\cite{Kawaguchi2012Spinor,Stamper2013Spinor,RevModPhys.90.035005}. This versatility has motivated spinor BECs as a basis for theoretical proposals and experimental studies of quantum phase transitions in both ground and excited states~\cite{PhysRevLett.124.043001,PhysRevA.104.L031305,Cabedo2021,Feldmann2021Interferometric,Meyer2023Excited,Chisholm2024}, the generation of large-scale many-particle entanglement~\cite{Zhang2013Generation,LiYouScience2017,Feldmann2018Interferometric,Pezze2019Heralded, RevModPhys.90.035005}, and in quantum-enhanced atomic clocks~\cite{Kruse2016Improvement} and interferometers~\cite{Hamley2012Spin,Anders2021Momentum}.

The high degree of controllability in spinor BECs also positions them as prime candidates for quantum simulations aimed at emulating the Hamiltonians of physically distinct systems. This approach has been extensively pursued for quantum simulations of condensed-matter phenomena with BECs~\cite{Jaksch2005,Bloch2012,Gross2017, Schafer2020}. Recently, simulations of quantum chemistry and molecular processes have gained prominence as promising directions in quantum simulation~\cite{PhysRevX.6.031007,PhysRevX.8.031022,CiracNature2019,Daley2022}.

Besides the ability to solve theoretically intractable models, the advantages of quantum simulations include precise observations of real-time dynamics on timescales that are easier to access than those of the original systems, along with a level of quantum control that would otherwise be unachievable. This way, predictions of the model can be explored experimentally across varied parameter ranges including with initial conditions that would be unattainable in the original context~\cite{Schneider2012}. This is illustrated by proposals for quantum simulations of molecular electron transfer reactions~\cite{Schlawin2020} and their recent implementations using trapped ions~\cite{So2024} and superconducting qubits~\cite{Devoret2024}.

In this work, we demonstrate that the Hamiltonian of spin-1 BECs can be exactly mapped onto the two-dimensional vibron model (2DVM). This model is widely used in theoretical studies of molecular vibrations and the characterization of molecular configurations, particularly the transition between linear and bent geometries~\cite{Iachello1995Algebraic,Iachello1996Algebraic,Perez2008Algebraic}. It has successfully reproduced a wide array of molecular spectroscopic data and has been instrumental in advancing theoretical understanding of these processes. 

We explore how spinor BEC states can be interpreted in terms of the configurations of triatomic molecules by employing the Wigner quasiprobability distribution in phase space. Triatomic molecules can adopt linear or bent structures, characterised by the position of the central atom (see Fig.~\ref{fig:coordinates}). The Wigner distribution provides insights into the position and momentum distribution of the central atom, thereby linking the quantum states of spinor BECs to molecular configurations and dynamics. Spinor BEC systems enable continuous variation of control parameters, flexible preparation of states of interest, and the observation of dynamical signatures of excited-state quantum phase transitions (ESQPTs), thereby offering access to a wide range of phenomena that cannot be studied directly in molecular systems.

We focus on two dynamical cases: bending a linear molecule or straightening a bent molecule. Particularly, we observe a dynamical instability in the latter case.
In the dynamics triggered by the corresponding instability, a significant amount of entanglement is generated, and we characterise the dynamics with the squeezing parameter and the quantum Fisher information (QFI). Non-Gaussian fluctuations, revealed by the difference between the squeezing parameter and the QFI, increase with the system size once the spinor system crosses into the bent phase. This gives rise to an experimentally observable, dynamical signature for the quantum phase transition.

This paper is structured as follows. We introduce the 2DVM in Sec.~\ref{sec:2DVM}, loosely following Ref.~\cite{Perez2008Algebraic}. In Sec.~\ref{sec:spinor} we review relevant properties of spinor BECs, their experimental controllability and flexibility. In Sec.~\ref{sec:preliminary}, we show how the molecular configuration is reflected in phase space through the Wigner quasiprobability distribution that informs us about the motion of the centre atom.  
In Sec.~\ref{sec:quench}, we study the dynamics generated when the Hamiltonian corresponds to a bent equilibrium geometry while the system is initially prepared in a linear configuration. 
In Sec.~\ref{sec:accessibility}, we discuss the experimental accessibility of our results.  
Finally, the conclusion is given in Sec.~\ref{sec:conclusions}.

\section{Two-dimensional vibron model} \label{sec:2DVM}

The algebraic approach has been highly successful in both nuclear and molecular physics, providing a framework to construct the Hamiltonian under study by using operators that belong to the dynamical or spectrum-generating algebra~\cite{Iachello1995Algebraic,Iachello1996Algebraic,Perez2008Algebraic,Iachello2015Lie,Khalouf2019Calculation,Khalouf2021Excited}.
They have proven to be a useful approximation for describing the Hilbert space of bound rotational and vibrational states in molecules~\cite{Iachello1995Algebraic}. Although these models are available in arbitrary dimensions~\cite{Cejnar2007Phase}, we focus on the case of two dimensions that is relevant for the linear-to-bent transitions of triatomic molecules~\cite{Iachello1996Algebraic}. Rotations and vibrations occur in a plane, and are linked to the appearance of a quantum phase transition in the ground state~\cite{Perez2008Algebraic} and in excited states~\cite{Caprio2008Excited,Larese2013Signatures,Cejnar2021Excited,Khalouf2022Anharmonicity,Khalouf2022Quantum,Khalouf2024Degeneracy,Larese2011study,Cejnar2008Impact}. In the following, we review the algebraic structure of this model; see also Refs.~\cite{Perez2008Algebraic,Iachello1995Algebraic,Iachello1996Algebraic,Larese2013Signatures} for a more detailed account.

\begin{figure}[bt]
    \centering
    \includegraphics[width=0.4\linewidth]{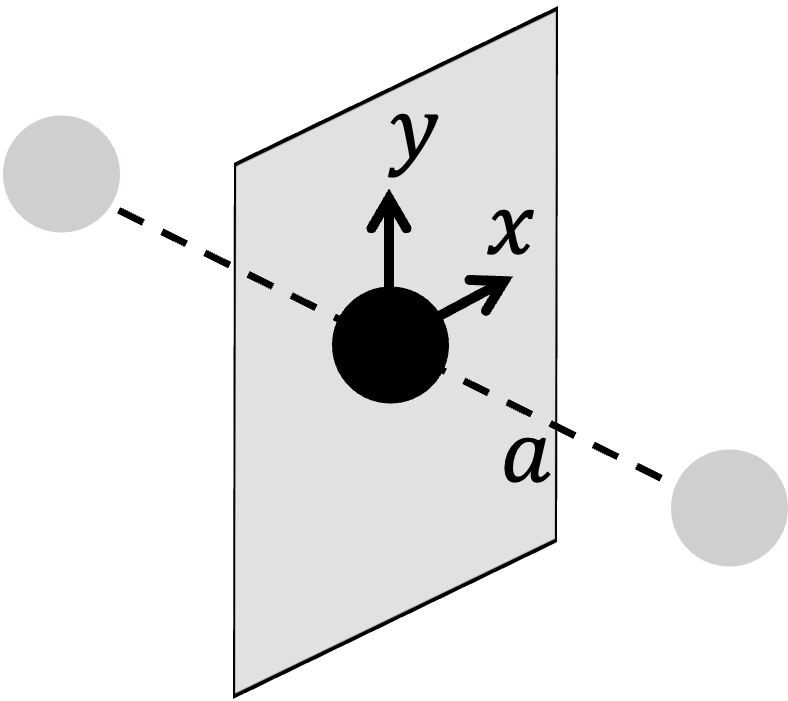}
    \caption{
    Coordinates of the centre atom reveal the configuration of a triatomic molecule.
    The bond length is denoted by $a$ and is the distance between neighboring atoms.
    }
    \label{fig:coordinates}
\end{figure}

The bosonic $U(3)$ Lie algebra, necessary for the description of two-dimensional systems, can be constructed using creation and annihilation operators for two Cartesian bosons $\{\hat{\tau}_x^\dagger, \hat{\tau}_y^\dagger, \hat{\tau}_x, \hat{\tau}_y\}$ and a scalar boson $\{ \hat{\sigma}^\dagger, \hat{\sigma} \}$ with the following commutation relations
\begin{subequations}
\begin{align}
    \left[\hat{\sigma}, \hat{\sigma}^\dagger \right] 
    &= 1,
    \\
    \left[\hat{\tau}_i, \hat{\tau}_j^\dagger \right] 
    &= 
    \delta_{ij},
    \\
    \left[\hat{\tau}_i, \hat{\sigma}^\dagger \right] 
    &= 0,
\end{align}
\end{subequations}
for $i,j = x,y$.
The bosonic operators $\hat{\tau}_{x,y}^{(\dagger)}$ express excitations in two dimensions, i.e. vibrons, and encode bending excitations, as will be explained later. 
For convenience, we introduce circular bosons, 
\begin{align} \label{eq:taup_taum}
    \hat{\tau}_{\pm}
    =
    \mp
    \frac{\hat{\tau}_{x}\mp i\hat{\tau}_{y}}{\sqrt{2}}
    .
\end{align}
The generators of the algebra $U(3)$ are the bilinear products of the creation and annihilation operators; for a full list see Eq.~\eqref{eq:3moleculealgebra} in Appendix~\ref{app:algebraic}. The algebraic approach to the construction of Hamiltonians consists in considering a linear combination of invariant Casimir operators with variable coefficients. Up to second order in the generators, the general Hamiltonian of two-dimensional problems then reads (see Appendix~\ref{app:algebraic} for details)
\begin{align} \label{eq:H2d}
    \hat{H}
    &=
    E_0 
    + \epsilon\hat{n}
    + \alpha \hat{n}\left(\hat{n}+1\right)
    + \beta \hat{l}^2
    + A \hat{W}^2
    ,
\end{align}
where $\epsilon,\alpha,\beta,A$ are free parameters that depend on the specific molecule and the operators used above are defined as
\begin{subequations}
\begin{align}
    \hat{n}
    &=
    \hat{\tau}_+^{\dagger}\hat{\tau}_+ + \hat{\tau}_-^{\dagger}\hat{\tau}_-,
    \\
    \hat{l}\label{eq:opl}
    &=
    \hat{\tau}_+^{\dagger}\hat{\tau}_+ - \hat{\tau}_-^{\dagger}\hat{\tau}_-,
    \\
    \hat{D}_+
    &=
    \sqrt{2}
    \left(
    \hat{\tau}_+^{\dagger}\hat{\sigma} - \hat{\sigma}^{\dagger}\hat{\tau}_-
    \right),
    \\
    \hat{D}_-
    &=
    \sqrt{2}
    \left(
    -\hat{\tau}_-^{\dagger}\hat{\sigma} + \hat{\sigma}^{\dagger}\hat{\tau}_+
    \right),
    \\
    \hat{W}^2
    &=
    \frac{1}{2}\left(
    \hat{D}_+\hat{D}_- + \hat{D}_-\hat{D}_+
    \right)
    +
    \hat{l}^2
    .
\end{align}
\end{subequations}
This is the most general two-body Hamiltonian for angular momentum-conserving, i.e., rotationally invariant systems.
This Hamiltonian interpolates between the two extreme cases of linear and bent molecular configurations, which are described by the respective Hamiltonians (see Appendix~\ref{app:algebraic_chainI} and~\ref{app:algebraic_chainII} for details)
% in Eqs.~\eqref{eq:chainI}\eqref{eq:chainII}. 
\begin{align}
    \hat{H}^{(I)}
    &=
    E_0 
    + \epsilon\hat{n}
    + \alpha \hat{n}\left(\hat{n}+1\right)
    + \beta \hat{l}^2
\end{align}
and
\begin{align}
    \hat{H}^{(II)}
    &=
    E_0 
    + \beta \hat{l}^2
    + A \hat{W}^2
    .
\end{align}
Whereas the general Hamiltonian~\eqref{eq:H2d} expresses the most general mixture of the two configurations, the essence of the quantum phase transition between them is already captured by only two non-commuting Casimir operators, leading to
\begin{align} \label{eq:H2d_essen}
    \hat H
    &=
    (1-\gamma)
    \hat{n}
    -
    \frac{\gamma}{N-1}
    \hat{W}^2
    ,
\end{align}
where $\gamma$ is a control parameter and we introduced the normalisation factor $1/(N-1)$ to account for two-body interactions in $\hat{W}$.
The molecular configurations are described by the following coherent state~\cite{Perez2008Algebraic},
\begin{align} \label{eq:cohernetstate}
    \ket{\phi}=
    \ket{x,y}
    &\equiv
    \frac{1}{\sqrt{N!}}
    \hat{b}_c^{\dagger N}\ket{0,0,0}
    ,
\end{align}
where $\ket{0,0,0}$ is a vacuum state and $\hat{b}_c^{\dagger}$ is the bosonic creation operator,
\begin{align} \label{eq:bcdagger}
    \hat{b}_c^{\dagger}
    &\equiv
    \frac{1}{\sqrt{1+x^2+y^2}}
    \left[
    \hat{\sigma}^{\dagger} + 
    \left( x\hat{\tau}_x^{\dagger} + y\hat{\tau}_y^{\dagger}
    \right)
    \right]
    .
\end{align}
This coherent state is called atomic coherent state~\cite{Radcliffe1971,PhysRevA.6.2211} or number-projected generalised coherent state of $U(3)$~\cite{Iachello1987The} which extends the idea of Glauber coherent states~\cite{PhysRev.131.2766} to systems with fixed particle number~\cite{MandelWolf}. 
The coefficients $x,y$ of the coherent state may be complex in general, but here, they will be restricted to real values. Importantly, the coefficients $x,y$ can be regarded as the average coordinates of the centre atom as depicted in Fig.~\ref{fig:coordinates}~\cite{Perez2008Algebraic}
and normalised by the bond length which is the distance between neighboring atoms.

In the limit $N\to\infty$, these coherent states can be used to study the mean-field limit of our model. This is particularly convenient to investigate the phase transition between the linear and the bent configurations. 
The energy density in the mean field limit $N\to\infty$ is expressed as
\begin{align} \label{eq:energydensity}
    \frac{\mel{x,y}{\hat{H}}{x,y}}{N}
    &=
    -(1-\gamma)\frac{1}{1+r^2}
    +
    \gamma
    \left(
    \frac{1-r^2}{1+r^2}
    \right)^2
    ,
\end{align}
where $r=\sqrt{x^2+y^2}$.
The minimum of this function depends on the control parameter $\gamma$. For $\gamma\leq\gamma_c\equiv1/5$, the minimum is at $r=0$, i.e., when the centre atom is situated on the axis that connects the two outer atoms. This situation describes a linear geometry. However, for $\gamma>\gamma_c$ the minimum appears at $r=r_0$, where
\begin{align} \label{eq:r0}
    r_0
    &\equiv
    \sqrt{\frac{5\gamma-1}{3\gamma+1}}
    .
\end{align}
This indicates that the ground-state configuration is a linear molecule for $\gamma\leq\gamma_c$ and molecules are bent for $\gamma>\gamma_c$, and the degree of bending is given by $r_0$, which depends on $\gamma$.
In other words, $r_0$ is the order parameter of the ground-state phase transition in this system.
The average number of excitations $\mel{\phi}{\hat{n}}{\phi}$, i.e., vibrons in the coherent state, bears witness to the phase transition as 
\begin{align}
    \mel{\phi}{\hat{n}}{\phi}
    &=
\begin{cases}
    0
    &
    \text{for } 
    \gamma \leq \gamma_c
    \\
    N\frac{r_0^2}{1+r_0^2}
    &
    \text{for } 
    \gamma > \gamma_c
    .
\end{cases}
\end{align}
This number increases sharply at $\gamma=\gamma_c$.

\begin{figure}[bt]
    \centering
    \includegraphics[width=0.7\linewidth]{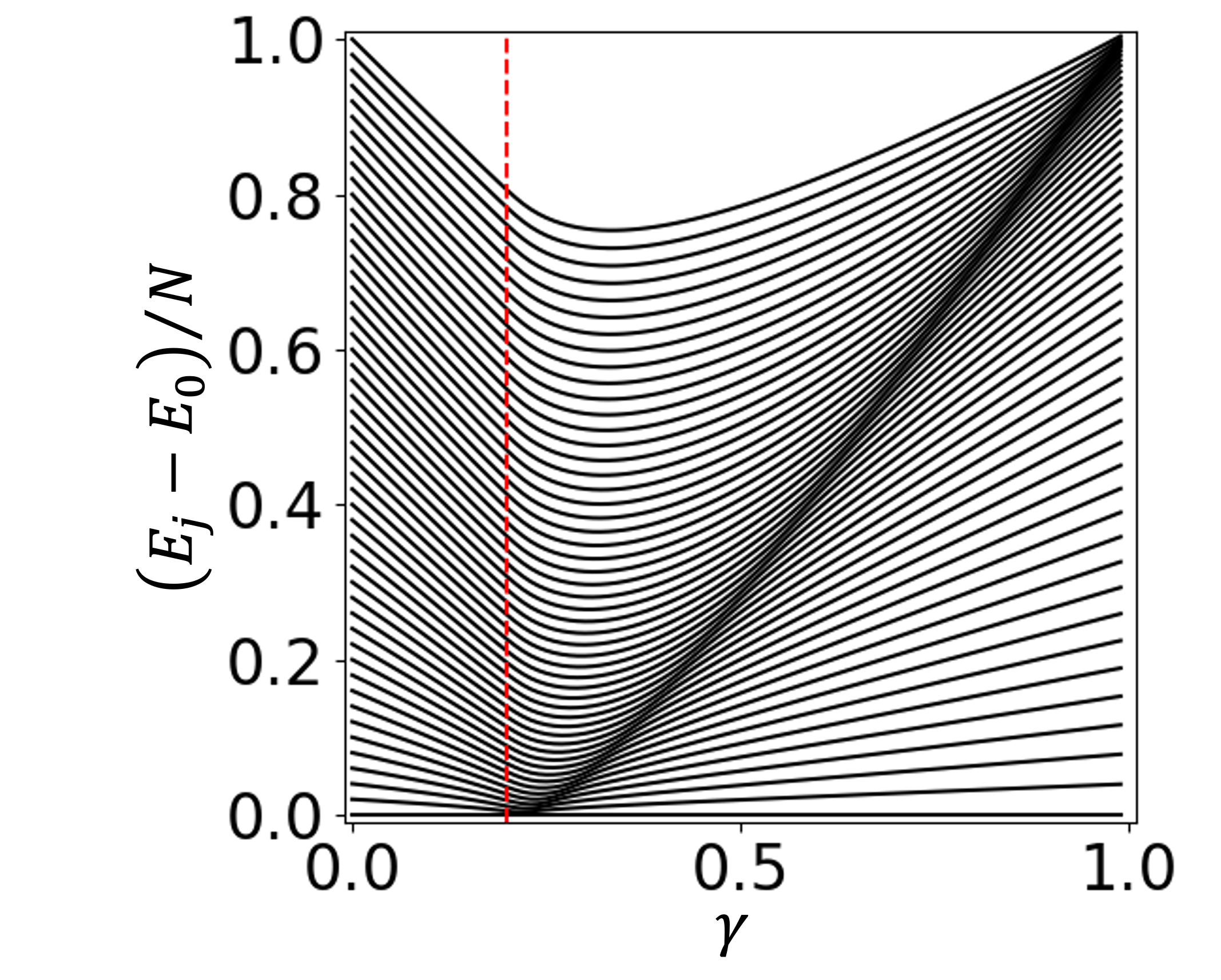}
    \caption{
    Energy spectrum of Eq.~\eqref{eq:H2d_essen} normalised by $N$ with zero magnetization for $N=100$ as a function of $\gamma$.
    The lowest energy is fixed to zero.
    The vertical dotted line highlights the position of $\gamma_c=1/5$.
    }
    \label{fig:energyspectrum_l0}
\end{figure}

In addition to the phase transition in the ground state, the 2DVM shows phase transitions in excited states~\cite{Caprio2008Excited,Larese2013Signatures}. 
We limit our attention to the subspace with zero magnetization, i.e., to states with eigenvalues zero for the vibrational angular momentum operator $\hat{l}$ in Eq.~\eqref{eq:opl}, and plot the energy of the Hamiltonian~\eqref{eq:H2d_essen} for $N=100$ as a function of $\gamma$ (see Fig.~\ref{fig:energyspectrum_l0}). Note that the lowest energy is fixed to zero. An avoided crossing in the two lowest energy states at $\gamma\approx\gamma_c$ indicates the ground state quantum phase transition. This avoided crossing also propagates over the excited states, and clusters of energy levels are seen for $\gamma>\gamma_c$ at a critical nonzero energy, characteristic of the ESQPT. The resulting divergence in the density of states splits the spectrum in Fig.~\ref{fig:energyspectrum_l0} in two regions. 
Left of the level clustering, the eigenstates have a linear character, whereas on the right they have a bent character~\cite{Larese2013Signatures}. 

The essential Hamiltonian~\eqref{eq:H2d_essen} of the 2DVM allows us to explore various configurations and the transition between them. However, the control parameter $\gamma$ is fixed by the molecule and cannot be tuned freely in a molecular setting. 
As we will discuss in the next section, spinor BECs can be operated in a regime that produces the same Hamiltonian with flexibly tunable parameters. 

\section{Spinor Bose-Einstein condensates} \label{sec:spinor}
We consider a finite-size spin-1 BEC in the $F = 1$ hyperfine ground state manifold in the presence of an external homogeneous magnetic field. The single-mode approximation~\cite{Kawaguchi2012Spinor,Stamper2013Spinor,Mirkhalaf2021Criticality} allows us to simplify the Hamiltonian for $N$ atoms, up to constant terms, to the following form
\begin{align}\label{eq:bec_spinor2}
    \hat{H}_{\text{spin-1}}
    &= 
    - q\hat{N}_0
    -\frac{c}{2N} \hat{J}^2.
\end{align}
Here, $\hat{J}^2 = \hat{J}_x^2 + \hat{J}_y^2 + \hat{J}_z^2$, where $\hat{J}_{x,y,z}$ are collective spin operators, and $\hat{N}_{m_F}$ is the number of atoms in the $m_F = 0, \pm1$ Zeeman state; see Appendix~\ref{app:spin1algebras} for additional details. The coefficient $q$ is the quadratic Zeeman energy shift.
The interaction coefficient $c$ is given by $c = N c_2 \int d\bf{r} \vert \phi(\bf{r}) \vert^4$ with $\phi(\bf{r})$ the spatial atomic wave function.
The coefficient $c_2$ is given by $c_2 = 4\pi \hbar^2 (a_0 - a_2)/3m$, where $m$ is the mass of each particle and $a_J$ is the s-wave scattering length for spin-1 atoms colliding in symmetric channels of total spin $J$. 
We consider $c_2$ positive which applies to ferromagnetic systems.

This system can be realised experimentally, for example with $\text{Rb}^{87}$ in the $F = 1$ hyperfine manifold. The coefficient $c_2$ can be tuned for example through changes of the potential that affect the spatial atomic wave function. The coefficient of the quadratic Zeeman term $q = q_B +q_{MW}$ depends on the external magnetic field $q_B$ and the microwave field $q_{MW}$. The part controlled by the magnetic field $B$ is $q_B = (\mu_BB)^2/4E_{HFS}$, where $\mu_B$ is Bohr magneton and $E_{HFS}$ is the hyperfine energy splitting. The value and the sign of $q_{MW}$ can be tuned independently of $q_B$ by employing a microwave field that is off-resonant with the other hyperfine state, allowing for flexible control of the value of $q$ in experiments that extends even into negative values~\cite{FeldmannThesis,Meyer2023Excited}.

To see how the Hamiltonian~\eqref{eq:bec_spinor2} of spinor BECs can be mapped onto the 2DVM Hamiltonian~\eqref{eq:H2d_essen}, we compare the following two operators in these models, 
\begin{subequations}
\begin{align}
    \hat{J}^2
    &=
    \hat{N}_0
    +\left(\hat{N}_{+1}-\hat{N}_{-1}\right)^2+2 \hat{N}_0\left(\hat{N}_{+1}+\hat{N}_{-1}\right)
    \nonumber\\
    &\quad
    +2 \hat{\tau}_{-}^{\dagger}\hat{\tau}_{+}^{\dagger}\hat{\sigma}\hat{\sigma}
    +2 \hat{\sigma}^{\dagger}\hat{\sigma}^{\dagger}\hat{\tau}_-\hat{\tau}_+
    ,
\end{align}
% Comparison with
\begin{align}
    \hat{W}^2
    &=
    \hat{N}_0
    +\left(\hat{N}_{+1}-\hat{N}_{-1}\right)^2+2 \hat{N}_0\left(\hat{N}_{+1}+\hat{N}_{-1}\right)
    \nonumber\\
    &\quad
    -2 \hat{\tau}_{-}^{\dagger}\hat{\tau}_{+}^{\dagger}\hat{\sigma}\hat{\sigma}
    -2 \hat{\sigma}^{\dagger}\hat{\sigma}^{\dagger}\hat{\tau}_-\hat{\tau}_+
    ,
\end{align}
\end{subequations}
where the identities hold up to constants proportional to the total particle number $N$.
This comparison reveals a close similarity except for the signs of the two final terms. 
These signs can be compensated by applying a $\pi/2$-rotation to mode $0$. 
In addition, by controlling the quadratic Zeeman term $q/c$ such that $q/c=2(1-\gamma)/\gamma$, the spinor BEC Hamiltonian~\eqref{eq:bec_spinor2} corresponds to the essential Hamiltonian~\eqref{eq:H2d_essen} up to a multiplicative factor,
\begin{align} \label{eq:H_spinor_2DVM}
    \frac{\hat{H}_{\text{spin-1}}}{c/2\gamma}
    &=
    -\left(1-\gamma\right) \hat{N}_0 
    -\frac{\gamma}{N} \hat{J}_{\text{rot}}^2
    , 
\end{align}
where $\hat{J}_{\text{rot}}^2\equiv\mathrm{e}^{i\hat{N}_0 \pi/2}\hat{J}^2\mathrm{e}^{-i\hat{N}_0 \pi/2}$ is the squared total spin operator in a rotating frame. 
The essential Hamiltonian~\eqref{eq:H2d_essen} for $\gamma=0$ is simply $-\hat{N}_0$ and, even though technically $\gamma$ in the spinor system~\eqref{eq:H_spinor_2DVM} is never exactly zero, this limit can be studied at early times where the weak interaction is negligible. 
To be consistent with the 2DVM Hamiltonian~\eqref{eq:H2d_essen}, we introduce $\hat{H}_{\text{spin-1}}'\equiv\hat{H}_{\text{spin-1}}(2\gamma/c)$ so that $\hat{H}_{\text{spin-1}}'$ matches the 2DVM Hamiltonian~\eqref{eq:H2d_essen} besides a multiplicative factor.
In the following, the time evolution generated by the Hamiltonian~\eqref{eq:H_spinor_2DVM} will be given in normalised time units $t/t_0$ with $t_0\equiv 2\gamma/c$ unless otherwise specified.

The correspondence between the two Hamiltonians~\eqref{eq:H2d_essen} and \eqref{eq:H_spinor_2DVM} implies that spinor BECs can act as analog simulators for the 2DVM. A broader class of bosonic two-level pairing models that includes the Lipkin-Meshkov-Glick (LMG) model, the Hamiltonian of spinor BECs, and the 2DVM are known to coincide under general conditions in the mean-field limit~\cite{Caprio2008Excited,FeldmannThesis}. As we have seen above, the exact mapping between spinor BECs and the 2DVM also holds beyond the mean-field approximation. 

Moreover, systems of spinor BECs provide high flexibility in the preparation of states as well as the controllability of the parameters. Experimental techniques analogous to homodyne detection have been developed for spinor BECs~\cite{Gross2011Atomic,Peise2015Satisfying}, and spin-coherent states~\eqref{eq:cohernetstate} can be prepared with high fidelity~\cite{RevModPhys.90.035005}. Also, the quantum phase transition in ground and excited states have been successfully studied~\cite{Kawaguchi2012Spinor,Bookjans2011Quantum,Zhang2013Generation,Zhang2005Coherent,Zibold2010Classical,Zhao2014Dynamics,FeldmannThesis,Meyer2023Excited}.

\section{Coherent molecular states in phase space}\label{sec:preliminary}

\begin{figure}[tb]
    \centering
    \includegraphics[width=0.7\linewidth]{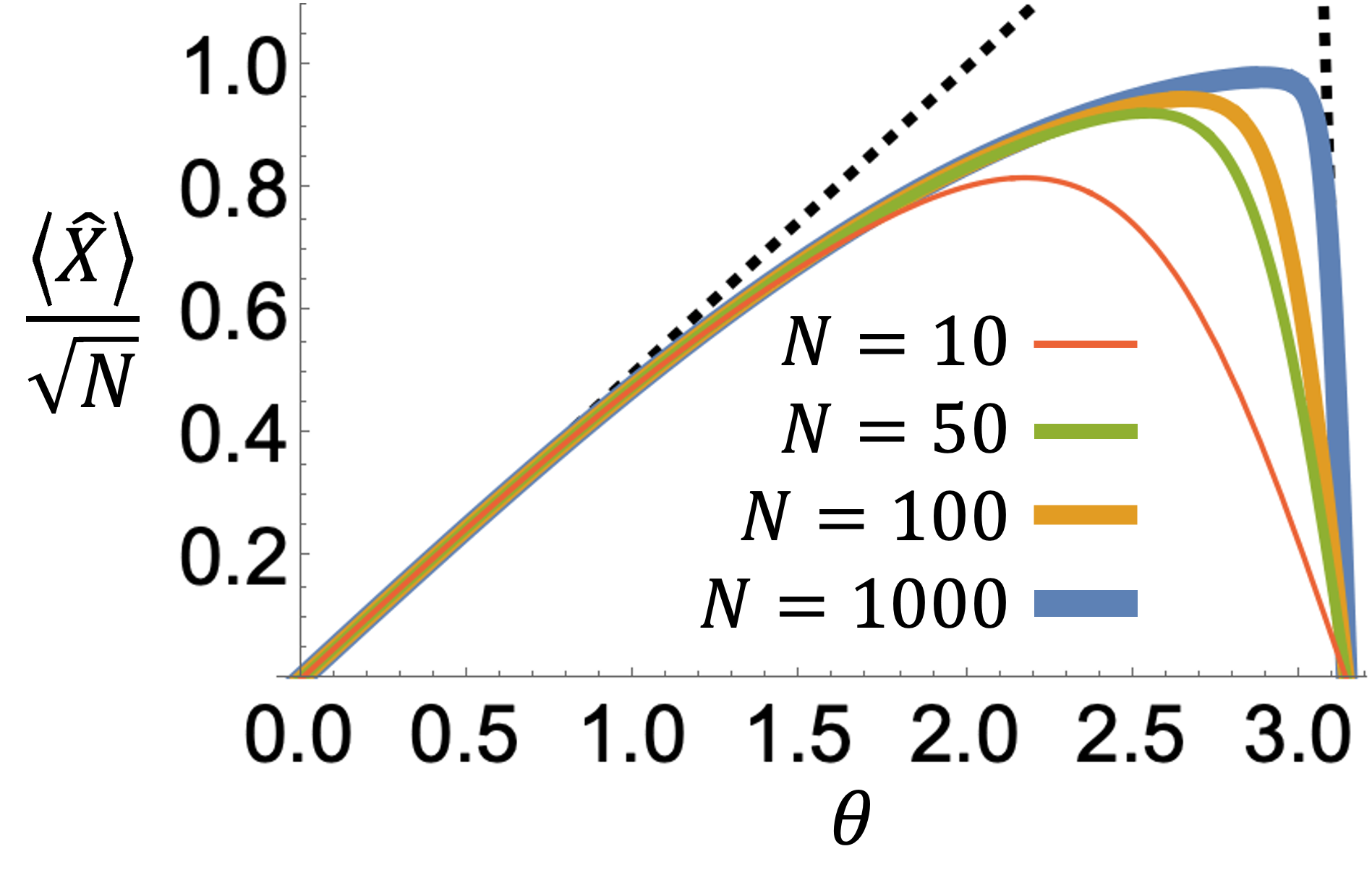}
    \caption{
    Displacement $\langle \hat{X} \rangle/\sqrt{N}$ of the coherent state~\eqref{eq:cohernetstate_1} for $N=10,50,100,1000$ as a function of $\theta$.
    The dotted lines are the asymptotes of $\langle \hat{X} \rangle/\sqrt{N}$ at $\theta=0,\pi$. The one going through $\theta=0$ is $\theta/2$, and the one going through $\theta=\pi$ is $\sqrt{N}(-\theta+\pi)/2$ for $N=1000$.
    }
    \label{fig:XP_coheretstate}
\end{figure}

The coherent state~\eqref{eq:cohernetstate} expresses the molecular configuration in phase space and reveals the relation between the coordinates $x,y$ of the centre atom.
For simplicity, we now focus on states with $y=0$, but it is straightforward to extend the discussion to the case of $x\neq 0$ and $y\neq 0$ by rotating $\hat{\tau}_{x,y}$ accordingly.
The coherent state~\eqref{eq:cohernetstate} with $y=0$ fixed has support in the subspace where only the modes $0,x$ are populated and can be rewritten as 
\begin{align} \label{eq:cohernetstate_1}
    \ket{\phi}=\ket{\theta}
    &=
    \sum_{n_x=0}^N
    \sqrt{\binom{N}{n_x}}
    \sin^{n_x}\frac{\theta}{2}
    \cos^{N-n_x}\frac{\theta}{2}
    \nonumber\\
    &\quad\quad\quad\quad\quad\quad\quad
    \ket{n_x,N-n_x,0}_{xy}
    ,
\end{align}
where the position $x\in(-\infty,\infty)$ is expressed by $\theta\in[0,\pi]$ through $x=\tan\frac{\theta}{2}$ and $\ket{n_x,n_0,n_y}_{xy}$ is the number state basis of the modes $x,0,y$. 
At $\theta=0,\pi$, this state corresponds to $\ket{0,N,0}_{xy}$ or $\ket{N,0,0}_{xy}$, respectively. To study the phase space of mode $x$ we introduce the quadratures $\hat{X}\equiv(\hat{\tau}_x+\hat{\tau}_x^{\dagger})/2$ and $\hat{P}_X\equiv(\hat{\tau}_x-\hat{\tau}_x^{\dagger})/2i$. 
The average displacement $\langle \hat{X} \rangle\equiv\mel{\phi}{\hat{X}}{\phi}$ of the coherent state~\eqref{eq:cohernetstate_1} is given by
\begin{align} \label{eq:X_coherent}
    \langle \hat{X} \rangle
    &=\left(\tan\frac{\theta}{2}\right)
    \sum_{k=0}^N
    c_k^2(\theta)
    \sqrt{N-k}
    ,
\end{align}
where $c_k^2(\theta)=\binom{N}{k}\sin^{2k}\frac{\theta}{2}\cos^{2(N-k)}\frac{\theta}{2}$ and $\sum_{k=0}^Nc_k^2(\theta)=1$.
Note that the coherent state $\ket{\phi}$ has vanishing average momentum, 
\begin{align} \label{eq:PX_coherent}
    \langle \hat{P}_X \rangle\equiv\mel{\phi}{\hat{P}_X}{\phi}
    &=
    0
    .
\end{align}
Figure~\ref{fig:XP_coheretstate} shows a plot of $\langle \hat{X} \rangle/\sqrt{N}$ as a function of $\theta$ for different $N$. The function $\langle \hat{X} \rangle/\sqrt{N}$ is concave and yields zero for $\theta=0,\pi$ for any $N$.
For small $\theta$, the function $\langle \hat{X} \rangle/\sqrt{N}$ is independent of $N$ since the state is close to the vacuum state $\ket{0}$. 
The position of the maximum approaches $\theta=\pi$ for larger $N$.
Furthermore, the asymptote at $\theta=\pi$ is $\sqrt{N}(-\theta+\pi)/2$ (the dashed line at $\theta=\pi$). Thus, the function is discontinuous at $\theta=\pi$ in the limit $N\to\infty$. 

\section{Quench dynamics of molecular bending in spinor BECs} \label{sec:quench}

\begin{figure*}[t]
    \centering
    \includegraphics[width=0.8\linewidth]{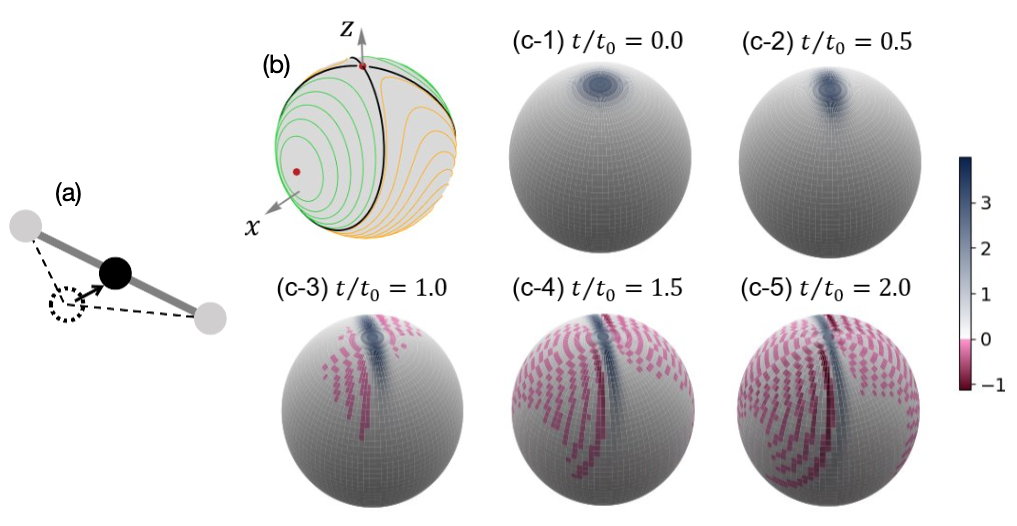}
    \caption{
    (a) Sketch of straightening a bent triatomic molecule. 
    (b) Trajectories on the mean-field model~\eqref{eq:spinormf} with fixed energy density $\eta$ for $\gamma=0.5$. The black curve represents the separatrix at $\eta=-(1-\gamma)$, and the green curves and the orange curves show the trajectories below and above the separatrix respectively. The red dots represent the stationary points (see text).
    (c) Time evolution of the Wigner quasiprobability distribution in mode $x$ subspace for $\gamma=0.5$ and $N=50$.
    }
    \label{fig:twomodesqueezing}
\end{figure*}

Here, we consider quenching such coherent states~\eqref{eq:cohernetstate_1} which express the molecular configurations in two cases: bending a linear molecule and straightening a bent molecule. The former case, where the system is prepared in a bent state and evolves under the linear Hamiltonian, can be regarded as a warm-up exercise and is presented in Appendix~\ref{app:linear2bent}. 
It shows a simple harmonic oscillation of the centre atom under the quench dynamics. In the present section, we focus on the latter case, where the Hamiltonian parameters correspond to a bent molecular configuration while the system is prepared initially in a linear configuration (see Fig.~\ref{fig:twomodesqueezing}(a)).
To this end, we prepare the coherent state~\eqref{eq:cohernetstate_1} for $\theta=0$, i.e. $\ket{0,N,0}_{xy}$, as the initial state and let it evolve under the Hamiltonian~\eqref{eq:H_spinor_2DVM} for $\gamma>\gamma_c$. Initially, only mode 0 is populated, and then particles are distributed to other modes through collisions. In the following we study the ensuing time evolution from the point of view of the bending vibration. 

In principle, it is possible to construct the Wigner quasiprobability distribution of three modes~\cite{Rowe1999Representations}, but this representation is difficult to analyse since it requires a four-dimensional surface. However, the essence of the dynamics can be extracted after a projection onto two-mode states, which are explained in the next subsection. 

We first remark that the low-depletion approximation that is often used in situations where most of the atomic population remains in the $m_F=0$ mode is unable to describe the phase transition between linear and bent configurations (see Appendix~\ref{app:lowdeplete} for details). Instead, as we will explain in the following, a suitable path towards exploring the phase transition is achieved by a mean-field approximation after projecting the spinor Hamiltonian~\eqref{eq:H_spinor_2DVM} onto the mode-$x$ subspace.

\subsection{Mean-field approximation in mode $x$ subspace}

Consider the basis $\ket{n_x,n_0,n_y}_{xy}$ of a three-mode system with a total particle number $N=n_x+n_0+n_y$.
For fixed $n_y$, the basis $\ket{n_x,n_0,n_y}_{xy}$ can be regarded as a basis $\ket{n_x,n_0}_x$ of a two-mode system with total particle number $N-n_y$. 
% Therefore, 
In this representation, a spin-1 system with $N$ particles can be decomposed into a collection of spin-$1/2$ systems with different total particle numbers $M=0,1,\ldots,N$ as
\begin{align} \label{eq:S1_S12}
    \mathcal{S}_1^{N}
    &=
    \mathcal{S}_{1/2}^{0}
    \oplus
    \mathcal{S}_{1/2}^{1}
    \oplus
    \cdots
    \oplus
    \mathcal{S}_{1/2}^{N}
    ,
\end{align}
where we denote these Fock-space sectors
% the set of the bases for a two-mode (three-mode) system 
with the total particle number $M$ as  $\mathcal{S}_{1/2}^{M}$ ($\mathcal{S}_{1}^{M}$).
We focus on the subspace $\mathcal{S}_{1/2}^{N}$ and project the system onto it.
Note that this subspace corresponds to the space LMG model expands, which is one of the bosonic two-level pairing models.

To observe the sharp features of the phase transition we consider the mean-field limit and study the structure of phase space in the $N\to\infty$ limit.
Similar analyses have been done in Refs.~\cite{Raggio1989Quantum,Duffield1992Mean,Duffield1992Classical} and particularly in a related model in a spinor system~\cite{Feldmann2021Interferometric,FeldmannThesis}.
Specifically, we determine the expectation value of the Hamiltonian~\eqref{eq:H_spinor_2DVM} on the spin-coherent state in mode $x$ subspace, defined as
\begin{align} \label{eq:spincoherent_sym}
    \ket{\tilde{\phi}}
    &=
    \ket{\varphi, \theta}
    \nonumber\\
    &\equiv
    \frac{1}{\sqrt{N!}}
    \left( 
    \hat{\tau}_x^\dagger {\sin}\frac{\theta}{2}
    + 
    \hat{\sigma}^\dagger {\cos }\frac{\theta}{2} e^{-i\varphi}\right)^N\ket{0,0}_x
    ,
\end{align}
where $\ket{n_x,n_0}_x$ is the number state basis of mode $x,0$ with $\varphi\in[0,2\pi)$ and $\theta\in[0,\pi]$. 
The angles $\varphi,\theta$ are set such that 
$\mel{\tilde{\phi}}{\hat{X}_{\rm x}}{\tilde{\phi}}
=(N/2)\sin\theta\cos\varphi
$, 
$\mel{\tilde{\phi}}{\hat{X}_{\rm y}}{\tilde{\phi}}
=(N/2)\sin\theta\sin\varphi$, 
and $\mel{\tilde{\phi}}{\hat{X}_{\rm z}}{\tilde{\phi}}
=(N/2)\cos\theta$. 
For $\varphi=0$, this spin-coherent state $\ket{\tilde{\phi}}$ corresponds to the spin-coherent state $\ket{\phi}$ defined in Eq.~\eqref{eq:cohernetstate_1} which describes the configuration of stationary molecules. 
By taking the limit $N\to\infty$, the energy density $h_{mf}^x\equiv\mel{\tilde{\phi}}{\hat{H}_{\text{spin-1}}'}{\tilde{\phi}}/N$ is given by
\begin{align}
    \label{eq:spinormf}
    h_{mf}^x
    &=
    -(1-\gamma)
    \frac{1}{2}
    \left(
    1+s
    \right)
    -
    \gamma
    \left(
    1-s^2
    \right)
    \cos^2 \varphi
    ,
\end{align}
where $s \equiv \cos\theta$ and $s\in[-1,1]$.
The maximum value of $h_{mf}^x$ is always 0 for $s=-1$, 
and the minimum value of $h_{mf}^x$ is given by the following three points: 
\begin{center}
\begin{tabular}{ c|c|c } 
 \hline
 $s$ & 1 & $\left(1/\gamma-1\right)/4$ \\
 $\cos\varphi$ & any & $\pm1$ \\
 \hline
 $\gamma$ & $0\leq\gamma\leq1/5$ & $1/5<\gamma\leq1$ \\ 
 \hline
 $h_{mf}^x$ & $-(1-\gamma)$ & $-(3\gamma+1)^2/16\gamma$ \\ 
 \hline
\end{tabular}
\end{center}
This is consistent with the previous result for the energy density~\eqref{eq:energydensity} that has different minima, depending on whether $\gamma$ is larger or smaller than $\gamma_c=1/5$~\cite{Perez2008Algebraic}.

The equations of motion for conjugate variables $( \varphi,s)$ are obtained as
\begin{subequations}
    \begin{align}
    &\dot{\varphi} 
    =
    2\gamma 
    s
    \cos^2\varphi
    -
    \frac{1-\gamma}{2}
    , 
    \\
    &\dot{s}
    = 
    2\gamma
    \left(
    1-s^2
    \right)
    \sin\varphi
    \cos\varphi
    . 
    \end{align}
\end{subequations}
The position of stationary points can be found by solving $(\dot{\varphi},\dot{s})=(0,0)$ and are given by the following table:
\begin{center}
\begin{tabular}{ c|c|c|c } 
 \hline
 $s$ & 1 & $\left(1/\gamma-1\right)/4$ & any \\
 $\cos\varphi$ & $\pm\sqrt{1/\gamma-1}/2$ & $\pm 1$ & $0$ \\
 \hline
 $\gamma$ & $1/5\leq \gamma\leq 1$ & $1/5\leq \gamma\leq 1$ & $\gamma=1$ \\ 
 \hline
\end{tabular}
\end{center}
While for $\gamma<1/5$ the energy density $h_{mf}^x$ does not have stationary points, for $1/5\leq\gamma\leq1$ three stationary points appear. 
One of the points is placed at $s=1$ and corresponds to the initial state $\ket{0,N}_x$.
The other two points are at the minimum of the energy density $h_{mf}^x$ and correspond to bent configurations. 

We have plotted the trajectories $(\varphi,s)$ that give level sets of $h_{mf}^x$ with energy density $\eta$ fixed for $\gamma=0.5$ in Fig.~\ref{fig:twomodesqueezing}(b). The black line is the trajectory at $\eta=-(1-\gamma)$ and divides two types of trajectories. It is called the separatrix and exists only for $\gamma>\gamma_c=1/5$. The three stationary points are marked with red dots, and the initial state is located at the intersection of the separatrix. 
Even though the initial state is stationary in the mean field limit, for finite $N$ the quantum fluctuations of the initial state extend over different types of classical trajectories. States on such different trajectories evolve in different directions, leading to a dynamical instability.

We stress that Eq.~(\ref{eq:spinormf}) is obtained after restricting the dynamics to the $x$ subspace and taking the mean-field limit at zero angular momentum. In this reduced sector, the effective classical Hamiltonian has the same structure as in LMG-type two-level bosonic models~\cite{FeldmannThesis}. The phase-space trajectories are properties of this reduced one-degree-of-freedom quench dynamics, rather than of the full 2DVM. Our main result, however, concerns the exact mapping between the spin-1 BEC and the essential 2DVM Hamiltonian beyond mean field, with the molecular interpretation inherited from that mapping, and it is expected that long-time dynamics of our Hamiltonian deviate from those of LMG-type two-level bosonic models, which will be studied in the next subsection.
Important differences between one- and higher-dimensional two-level bosonic models have been discussed in Ref.~\cite{Khalouf2024Degeneracy}.

% \subsection{Exact dynamics in mode $x$ subspace}
\subsection{Exact dynamics beyond mode $x$ subspace}

Based on the result of the mean-field model, we move on to the exact time evolution under the full Hamiltonian~\eqref{eq:H_spinor_2DVM}, i.e. beyond the reduced sector $\mathcal{S}_{1/2}^{N}$.
For simplicity, we restrict to the zero-magnetization subspace, i.e., $\langle\hat{J}_z\rangle=0$ and $(\Delta\hat{J}_z)^2=0$, where the variance for any operator $\hat{O}$ is defined as $(\Delta\hat{O})^2\equiv\langle\hat{O}^2\rangle-\langle\hat{O}\rangle^2$. This is due to the fact that magnetization $\hat{J}_z$ is conserved because of $[\hat{J}_z,\hat{H}_{\text{spin-1}}']=0$ and that the magnetization of the initial state vanishes.
We compute the exact time evolution and project it onto mode $x$ subspace to construct the Wigner quasiprobability distribution on the Bloch sphere for $\gamma=0.5$ and $N=50$ (see Fig.~\ref{fig:twomodesqueezing}(c)).
% study the Wigner quasiprobability distribution on the Bloch sphere in mode $x$ subspace for $\gamma=0.5$ and $N=50$ (see Fig.~\ref{fig:twomodesqueezing}(c)).
The distribution expands along the separatrix, becomes squeezed, and shows negativity in certain regions. 

To characterise the squeezing and the creation of the entanglement, we employ the squeezing parameter, which is defined as~\cite{PhysRevA.50.67,Wineland1992Spin,RevModPhys.90.035005}
\begin{align}
    \xi^2
    &=
    \frac{N\left(\Delta \hat{X}_{n_3}\right)^2}{\langle \hat{X}_{n_1}\rangle^2}
    ,
\end{align}
and the quantum Fisher information (QFI)~\cite{Helstrom1969Quantum,Braunstein1994Statistical}, which for pure states is given by
\begin{align}
    F_{Q}\left[
    \hat{X}_{n_2}
    \right]
    &=
    4
    \left(
    \Delta \hat{X}_{n_2}
    \right)^2
    ,
\end{align}
where $\hat{X}_{n_j}$ is the spin operator along the direction $n_j$ in mode $x$ subspace (see Eq.~\eqref{eq:SU2_modex} in Appendix~\ref{app:spin1algebras}) and $n_1$, $n_2$, $n_3$ are three mutually orthogonal directions. 
Sufficient criteria for entanglement are given by~\cite{Pezze2009Entanglement,RevModPhys.90.035005}
\begin{align}\label{eq:spinsqz}
    \xi^2<1,
\end{align}
and
\begin{align}\label{eq:QFIent}
    \zeta^2
    &=
    \frac{N}{F_{Q}\left[
    \hat{X}_{n_2}
    \right]}
    <1
    .
\end{align}
Spin-squeezed states, which satisfy Eq.~(\ref{eq:spinsqz}), exhibit quantum fluctuations below that of a collection of $N$ identically prepared independent spin systems, while at the same time maintaining a large mean spin length oriented in a direction perpendicular to these fluctuations.
The condition~(\ref{eq:QFIent}) is satisfied by a wider class of metrologically useful entangled states and this condition follows from the convexity and the additivity of QFI of separable states, together with the fluctuation limit for a single spin $4\left(\Delta \hat{\sigma}_{n_2}\right)^2\leq1$~\cite{Pezze2014Quantum}.

Furthermore, the inequality $\zeta^2\leq\xi^2$ holds as the QFI captures the full metrological sensitivity of a quantum state, while spin squeezing can be seen as a Gaussian approximation~\cite{Pezze2009Entanglement,Gessner2019}. 
The maximum value of the QFI, $F_{Q}\left[\hat{X}_{n_2}\right]$, is $N^2$, and thus $1/N\leq\zeta^2\leq\xi^2$~\cite{Pezze2009Entanglement}.
To detect entanglement thoroughly, we optimise the directions $\{n_1,n_2,n_3\}$ such that $\xi^2$ and $\zeta^2$ are minimised.
The initial state occupies only mode 0 and thus maximises $\hat{X}_z$, i.e., $\langle\hat{X}_z\rangle=N/2$, and the expectation values of other directions remain invariant under the dynamics, $\langle\hat{X}_x\rangle=0$ and $\langle\hat{X}_y\rangle=0$, for all the times. Therefore, we take $n_1=z$ and choose $n_2,n_3$ on the $xy$ plane such that the variance  $\left(\Delta X_{n_i}\right)^2$ is maximised or minimised. This can be done by using the eigenvalues of the renormalised covariance matrix 
$\Sigma_{i,j}=2(\langle\hat{X}_i\hat{X}_j+\hat{X}_j\hat{X}_i\rangle-2\langle\hat{X}_i\rangle\langle\hat{X}_j\rangle)/N$, where $\langle\hat{X}_{i}\rangle=0$ for all $i$ in our case.
Denoting the eigenvalues of $\Sigma_{i,j}$ with $\lambda_{\pm}$, the optimal squeezing parameter and the optimal inverse QFI are given by~\cite{Sorelli2019Fast}
\begin{align} \label{eq:xi2_opt}
    \xi_{\text{opt}}^2
    &=
    \frac{N^2\lambda_-}{4\langle \hat{X}_{z}\rangle^2}
    ,
\end{align}
and
\begin{align} \label{eq:zeta2_opt}
    \zeta_{\text{opt}}^2
    &=
    \frac{1}{\lambda_+}
    .
\end{align}

\begin{figure}[t]
    \centering
    \includegraphics[width=0.99\linewidth]{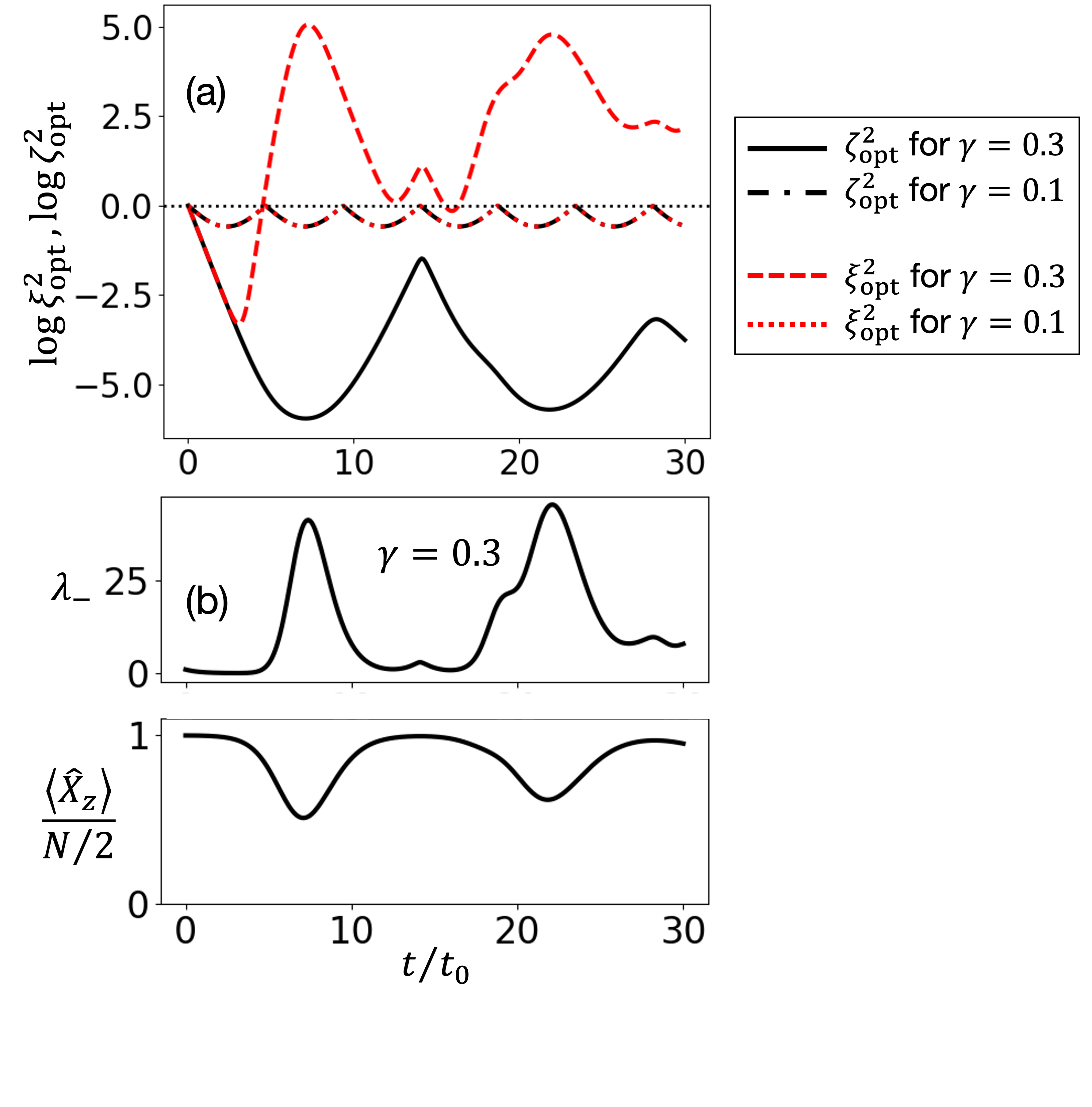}
    \vspace{-15mm}
    \caption{
    (a) Entanglement criteria $\xi_{\text{opt}}^2$~\eqref{eq:xi2_opt} and $\zeta_{\text{opt}}^2$~\eqref{eq:zeta2_opt} over time for $N=1000$ and $\gamma=0.1,0.3$ in the logarithmic scale.
    Values below the classical limit ($\log 1=0$) given by the dotted line indicate the presence of entanglement.  
    (b) Time evolution of the components of the optimal squeezing $\xi_{\text{opt}}^2$, $\lambda_-$ and $\langle\hat{X}_z\rangle/(N/2)$ for $\gamma=0.3$.
    }
    \label{fig:squeezing_QFI_time}
\end{figure}

The entanglement criteria $\xi_{\text{opt}}^2$, $\zeta_{\text{opt}}^2$ are plotted for $\gamma=0.1$ and $\gamma=0.3$ in Fig.~\ref{fig:squeezing_QFI_time}(a).
For $\gamma=0.1<\gamma_c$, the optimal squeezing parameter $\xi_{\text{opt}}^2$ and the optimal inverse QFI $\zeta_{\text{opt}}^2$ both drop below the classical limit of 1 and match each other closely over a long time span, indicating the presence of Gaussian entanglement. 
On the other hand, for $\gamma=0.3>\gamma_c$ the optimal inverse QFI $\zeta_{\text{opt}}^2$ becomes smaller than the optimal squeezing parameter $\xi_{\text{opt}}^2$ at some point. This indicates the non-Gaussian nature of the time-evolved state. Moreover, the stronger violation of the classical bound indicates the presence of multipartite entanglement among larger sets of spins~\cite{Hyllus2012Fisher,PhysRevA.85.022321,Ren2021}.

\begin{figure}[t]
    \centering
    \includegraphics[width=0.95\linewidth]{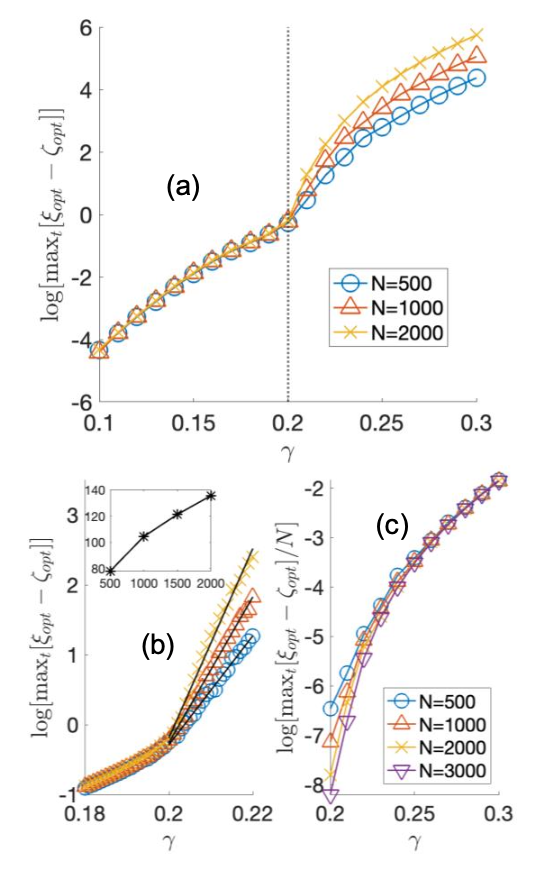}
    \caption{
    (a,b) Maximal difference $\xi_{\text{opt}}^2-\zeta_{\text{opt}}^2$ between criteria of entanglement detection~Eqs.~(\ref{eq:xi2_opt}) and \eqref{eq:zeta2_opt} over time for $N=500,1000,2000$ (see text). The time frame taken is $t\in[0,1000]$ with the number of points $10000$.
    In (b), the range of $\gamma$ is zoomed into $[0.18,0.22]$, and linear fittings for $\gamma>\gamma_c$ are added. The inset shows the slopes of the fittings as a function of $N$. 
    (c) Maximal difference $\xi_{\text{opt}}^2-\zeta_{\text{opt}}^2$ divided by $N$ for $N=500,1000,2000,3000$ for $\gamma\in[0.2,0.3]$.
    Notice that the vertical axis is the logarithmic scale.
    }
    \label{fig:diff_squeezing_QFI_gamma}
\end{figure}

Now, we turn our attention to the fact that, for $\gamma=0.3>\gamma_c$, the entanglement criteria $\zeta_{\text{opt}}^2$, $\xi_{\text{opt}}^2$ take extreme values at some points of time such as $t/t_0\sim 7$.
We note that the reason why the optimal squeezing parameter $\xi_{\text{opt}}^2$ reaches a peak is not just that the expectation value $\langle\hat{X}_z\rangle$ shrinks but also that the eigenvalue $\lambda_-$ of the covariance matrix $\Sigma_{i,j}$ grows (see Fig.~\ref{fig:squeezing_QFI_time}(b)).
In addition, the optimal inverse QFI $\zeta_{\text{opt}}^2$ keeps declining. A local maximum of the difference $\xi_{\text{opt}}^2-\zeta_{\text{opt}}^2$ arises when $\xi_{\text{opt}}^2$ is maximal and $\zeta_{\text{opt}}^2$ is minimal, and such local maxima of the difference appear periodically.
This is also observed for other values of $\gamma>\gamma_c$.
We have plotted the maximal difference $\max_t[\xi_{\text{opt}}^2-\zeta_{\text{opt}}^2]$ as a function of $\gamma$ for different $N$ (see Fig.~\ref{fig:diff_squeezing_QFI_gamma}(a)). 
The scaling changes at $\gamma=\gamma_c=0.2$ (see panel~(b)). 
This is because for $\gamma<\gamma_c$ the two quantities $\xi_{\text{opt}}^2$ and $\zeta_{\text{opt}}^2$ remain comparable, whereas for $\gamma>\gamma_c$ the contribution of $\xi_{\text{opt}}^2$ dominates the difference. This leads to different scalings below and above $\gamma=\gamma_c$.
In the region of $\gamma<\gamma_c$, the maximal differences for different $N$ coincide (see panel~(a)).
On the other hand, for $\gamma>\gamma_c$, the maximal differences grow with $N$, and after a transient regime close to the critical point $\gamma_c$ the scaling becomes linear in $N$ (see panel~(c)). 
These results, shown in Fig.~\ref{fig:diff_squeezing_QFI_gamma}, imply a non-smooth behavior of $\max_t[\xi_{\text{opt}}^2-\zeta_{\text{opt}}^2]/N$ in the thermodynamic limit $N\to\infty$, revealing the underlying quantum phase transition. 
 
\subsection{Interpretation of the dynamics in terms of bending vibration}

Up to now, we have projected the system to mode $x$ subspace to study trajectories with fixed energy in the mean field limit and analyse the dynamical instability. 
Here, again we focus on the same subspace $\mathcal{S}_{1/2}^{N}$ which is spanned by the basis $\{\ket{n_x,N-n_x,0}_{xy}\}$, and relabel the basis $\ket{n_x,N-n_x,0}_{xy}$ to the single mode basis $\ket{n_x}$.
This enables us to work on phase space $(X,P_X)$ and provides a more visual connection with the behaviour of a triatomic molecule. 
To this end, we transform the trajectories and the Wigner quasiprobability distribution on the Bloch sphere to phase space. 

First, we map the coherent state~\eqref{eq:spincoherent_sym} to phase space $(X,P_X)$ by binomially expanding Eq.~\eqref{eq:spincoherent_sym},
\begin{align} \label{eq:coherentstate_tilde}
    \ket{\tilde{\phi}}
    &=
    \sum_{n_x=0}^N
    \sqrt{\binom{N}{n_x}}
    \sin^{n_x}\frac{\theta}{2}
    \cos^{N-n_x}\frac{\theta}{2} 
    \nonumber\\
    &\quad\quad\quad\quad
    e^{-i(N-n_x)\varphi}
    \ket{n_x}
    ,
\end{align}
which corresponds to Eq.~\eqref{eq:cohernetstate_1} for $\varphi=0$.
The Cartesian coordinates of the coherent state on phase space are given by 
\begin{subequations} \label{eq:XP_spin_coherent}
\begin{align}
    \mel{\tilde{\phi}}{\hat{X}}{\tilde{\phi}}
    &=
    \cos\varphi
    \left(
    \tan\frac{\theta}{2}
    \sum_{k=0}^{N}
    \sqrt{N-k}
    c_k^2(\theta)
    \right)
    ,
\end{align}
and
\begin{align}
    \mel{\tilde{\phi}}{\hat{P}_X}{\tilde{\phi}}
    &=
    \sin\varphi
    \left(
    \tan\frac{\theta}{2}
    \sum_{k=0}^{N}
    \sqrt{N-k}
    c_k^2(\theta)
    \right)
    .
\end{align}
\end{subequations}
Equations~\eqref{eq:XP_spin_coherent} inform us about the average position and momentum of the state~\eqref{eq:coherentstate_tilde} in phase space. The phase $\varphi$ indicates the azimuth on phase space, and the function $\tan\frac{\theta}{2}\sum_{k=0}^{N}\sqrt{N-k}c_k^2(\theta)$ can be regarded as the radius of phase space. This radius corresponds to Eq.~\eqref{eq:X_coherent}, and thus the relation between the radius and $N,\gamma$ is displayed in Fig.~\ref{fig:XP_coheretstate}.
Note that Eqs.~\eqref{eq:XP_spin_coherent} coincide with Eqs.~\eqref{eq:X_coherent} and~\eqref{eq:PX_coherent} if $\varphi=0$, and thus the coherent state~\eqref{eq:coherentstate_tilde} is an extension of Eq.~\eqref{eq:cohernetstate_1} with non-zero momentum.

\begin{figure}[t]
    \centering
    \includegraphics[width=0.99\linewidth]{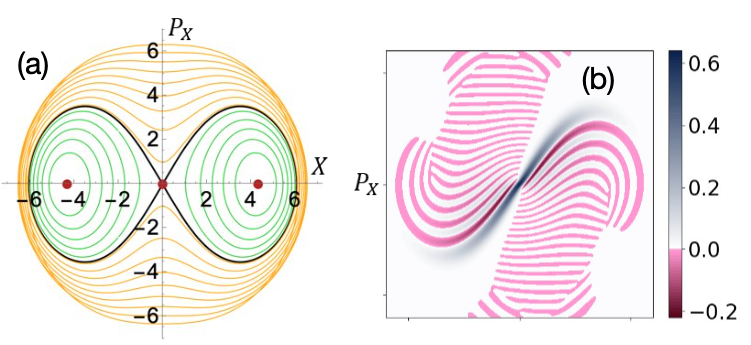}
    \caption{
    (a) Trajectories on the mean-field model with fixed energy $\eta$ mapped to phase space $(X,P_X)$ for $N=50$ and $\gamma=0.5$. 
    (b) Time evolution of the Wigner quasiprobability distribution in phase space $(X,P_X)$ for $N=50$.
    }
    \label{fig:mapping_twomode_XP}
\end{figure}

One can convert the set of trajectories with fixed energy on the Bloch sphere in mode $x$ subspace depicted in Fig.~\ref{fig:twomodesqueezing}(b) to the phase space $(X,P_X)$ by using Eqs.~\eqref{eq:XP_spin_coherent}. A caveat is that the radius of Eqs.~\eqref{eq:XP_spin_coherent} is dependent on $N$ while the set of trajectories are obtained in the limit $N\to\infty$. However, the radius is insensitive to $N$ in the region of $\theta\lesssim\theta_{\text{max}}$ for large $N$ with $\theta_{\text{max}}$ the maximum point according to Fig.~\ref{fig:XP_coheretstate}. Thus, the conversion of the trajectories poses no issues up to $\theta\lesssim\theta_{\text{max}}$ for large $N$, and Fig.~\ref{fig:mapping_twomode_XP}(a) displays the trajectories on phase space for $\gamma=0.5$ and $N=50$.
The three red dots represent the stationary points. One of them is located at the origin $(X,P_X)=(0,0)$ and indicates a linear molecule. The other two points are on the $X$ axis and correspond to bent molecules. As expected, the figure shows that the dynamics are unstable (stable) if the initial state corresponds to a linear molecule (a bent molecule). 
If the initial state is close to the linear molecule but has non-vanishing momentum or is even slightly bent, the centre atom shows trivial rotation that follows the orange or green trajectories in the figure. 

Furthermore, we have computed the Wigner quasiprobability distribution on phase space $(X,P_X)$ (see Fig.~\ref{fig:mapping_twomode_XP}(b)), and it provides a phase-space representation in terms of the position and momentum of the centre atom.
The probability develops symmetrically and becomes squeezed along the separatrix, leading to entanglement generation among vibrons. 
Recall that the molecule configuration is interpreted from a classical point of view according to the description of bent and linear molecules by coherent states given by Eqs.~\eqref{eq:cohernetstate_1} and~\eqref{eq:coherentstate_tilde}. Once the state considered deviates far from the coherent state, it is not straightforward to describe the molecule configuration, and rather one needs to employ vibrons, quantised excitations in two dimensions.
Nevertheless, since the Gaussian approximation is valid at the beginning of the dynamics, this squeezing can be interpreted as the emergence of a bent configuration with non-zero momentum, growing along with the separatrix.
At later times, the time-evolved state becomes too strongly non-Gaussian to employ this simple interpretation. This is further underscored by the appearance of negativity of the Wigner quasiprobability that cannot be understood classically. 
We note that dynamical consequences of ESQPTs in Wigner quasiprobability distributions in phase space have also been investigated in a squeeze-driven Kerr oscillator~\cite{Chavez2023Spectral}.

\section{Experimental accessibility}\label{sec:accessibility}

Finally, we comment on the experimental accessibility of the results we have obtained. The entanglement criteria used to characterise the quench dynamics are expressed in terms of expectation values and variances of spin operators and are commonly studied in experimental systems~\cite{RevModPhys.90.035005}. For realistic systems, described by mixed states, measurements of the QFI cannot be done directly from the variance; nevertheless, a variety of methods are available to access this property in atomic systems~\cite{Strobel2014Fisher}.

We used the Wigner quasiprobability in the dominant subspace to simulate the centre atom of a triatomic molecule. However, the main features of the molecular configuration are already observable from the average values of the quadratures, $\langle\hat{X}\rangle, \langle\hat{P}_X\rangle$, that can be readily accessed in atomic systems with techniques analogous to homodyne detection~\cite{Peise2015Satisfying,Gross2011Atomic}. Even the reconstruction of a Wigner quasiprobability from experimental data is feasible~\cite{Riedel2010Atom,McConnell2015Entanglement}. With a collection of Wigner quasiprobabilities in different subspaces given in Eq.~\eqref{eq:S1_S12}, it is even possible to build the full Wigner quasiprobability of a spinor BEC~\cite{Rowe1999Representations}.

Furthermore, we have shown the dynamical detection of the ground state phase transition, but also a method to detect the ESQPT in a spinor system has been proposed and demonstrated~\cite{Feldmann2021Interferometric,Meyer2023Excited}.
Such experimental flexibility makes spinor BECs a promising simulator platform, and our results have shown that spinor BECs can pave the way toward a deeper understanding of the dynamics of triatomic molecules.

\section{Conclusions} \label{sec:conclusions}

We have shown how spinor BECs can be employed as analog simulators of the 2DVM for polyatomic molecules. 
In contrast to actual molecular systems, spinor BECs allow one to change the control parameter continuously, prepare states of interest, and observe dynamical signatures such as squeezing. 
Using numerical simulations of BEC systems, we have studied the quantum dynamics associated with the quantum phase transition from linear to bent molecular configurations. Information about the molecular configuration is reflected in the atomic Wigner quasiprobability distribution, which contains information about quadrature observables. We have found that the dynamical instability associated with the evolution of a linear molecule in a bent potential leads to the production of a significant amount of atomic spin entanglement that can be captured with metrological tools such as the QFI and its Gaussian approximation, the squeezing parameter. Remarkably, the crossing of the critical parameter is heralded by an abrupt increase of the non-Gaussian metrological entanglement. The change in scaling with the system size leads to an observable witness of the bending transition in the atomic system.

Our work opens up a path towards exploring the quantum dynamics of molecular bending dynamics under controlled conditions with access to nonclassical quantum states. This flexibility, combined with tools from quantum information science, allows us to study the production of entanglement, squeezing, and to pinpoint the quantum phase transition in a quantum simulator.

\section{Acknowledgments}

AU and AN have contributed equally to this work.
AU is financially supported by JSPS Overseas Research Fellowships and Maria de Maeztu project (Grant CEX2019-000918-M funded by MCIN/AEI/10.13039/501100011033) and acknowledges further support from the Agencia Estatal de Investigaci\'{o}n and the Ministerio de Ciencia e Innovaci\'{o}n.
AU also acknowledges financial support from Spanish MCIN (MCIN/AEI/10.13039/501100011033, contract No. PID2022-141283NB-I00 and No. PID2022-139099NB-I00).
with the support of FEDER funds.
AN was supported by Government of Spain (Severo Ochoa CEX2019-000910-S, TRANQI, European Union NextGenerationEU PRTR-C17.I1), European Union (PASQuanS2.1, 101113690) and the ERC AdG CERQUTE.
This work was supported by the project PID2023-152724NA-I00, with funding from MCIU/AEI/10.13039/501100011033 and FSE+ and by the project CNS2024-154818, with funding from MICIU/AEI/10.13039/501100011033. This work was funded by MCIN/AEI/10.13039/501100011033 and the European Union ‘NextGenerationEU’ PRTR fund [RYC2021-031094-I], by the Ministry of Economic Affairs and Digital Transformation of the Spanish Government through the QUANTUM ENIA Project call—QUANTUM SPAIN Project, by the European Union through the Recovery, Transformation and Resilience Plan—NextGenerationEU within the framework of the Digital Spain 2026 Agenda, and by the CSIC Interdisciplinary Thematic Platform (PTI+) on Quantum Technologies (PTI-QTEP+). This work was supported by the Generalitat Valenciana through the project CIPROM/2022/066. This work was supported through the project CEX2023-001292-S funded by MCIU/AEI.\\

\section{Data Availability}

We have employed QuTiP for numerical calculations in this work~\cite{QuTiP,QuTiP2}, and our code is available online~\cite{Usui2025Code}.

\section{Author Contributions}

MG conceived the core theoretical idea and planned the study. 
AU and AN developed the analytical framework and performed the numerical simulations. 
All authors discussed the results, analyzed the data, and contributed to writing and revising the manuscript. 
Any AI models were not used. 

\bibliographystyle{apsrev4-1}
\bibliography{bibliografia}

@article{Perez2008Algebraic,
  title = {Algebraic approach to two-dimensional systems: Shape phase transitions, monodromy, and thermodynamic quantities},
  author = {P\'erez-Bernal, F. and Iachello, F.},
  journal = {Phys. Rev. A},
  volume = {77},
  issue = {3},
  pages = {032115},
  numpages = {21},
  year = {2008},
  month = {Mar},
  publisher = {American Physical Society},
  doi = {10.1103/PhysRevA.77.032115},
  url = {https://link.aps.org/doi/10.1103/PhysRevA.77.032115}
}

@article{Gross2011Atomic,
	author = {Gross, C. and Strobel, H. and Nicklas, E. and Zibold, T. and Bar-Gill, N. and Kurizki, G. and Oberthaler, M. K.},
	date = {2011/12/01},
	id = {Gross2011},
	isbn = {1476-4687},
	journal = {Nature},
	number = {7376},
	pages = {219--223},
	title = {Atomic homodyne detection of continuous-variable entangled twin-atom states},
	url = {https://doi.org/10.1038/nature10654},
	volume = {480},
	year = {2011}
}

@article{Peise2015Satisfying,
	author = {Peise, J. and Kruse, I. and Lange, K. and L{\"u}cke, B. and Pezz{\`e}, L. and Arlt, J. and Ertmer, W. and Hammerer, K. and Santos, L. and Smerzi, A. and Klempt, C.},
	date = {2015/11/27},
	doi = {10.1038/ncomms9984},
	id = {Peise2015},
	isbn = {2041-1723},
	journal = {Nature Communications},
	number = {1},
	pages = {8984},
	title = {Satisfying the Einstein--Podolsky--Rosen criterion with massive particles},
	url = {https://doi.org/10.1038/ncomms9984},
	volume = {6},
	year = {2015}
}

@article{Rowe1999Representations,
    author = {Rowe, D. J. and Sanders, B. C. and de Guise, H.},
    title = "{Representations of the Weyl group and Wigner functions for SU(3)}",
    journal = {Journal of Mathematical Physics},
    volume = {40},
    number = {7},
    pages = {3604-3615},
    year = {1999},
    month = {07},
    issn = {0022-2488},
    doi = {10.1063/1.532911}
}

@article{QuTiP,
title = {QuTiP: An open-source Python framework for the dynamics of open quantum systems},
journal = {Computer Physics Communications},
volume = {183},
number = {8},
pages = {1760-1772},
year = {2012},
issn = {0010-4655},
doi = {https://doi.org/10.1016/j.cpc.2012.02.021},
url = {https://www.sciencedirect.com/science/article/pii/S0010465512000835},
author = {J.R. Johansson and P.D. Nation and Franco Nori}
}

@article{QuTiP2,
title = {QuTiP 2: A Python framework for the dynamics of open quantum systems},
journal = {Computer Physics Communications},
volume = {184},
number = {4},
pages = {1234-1240},
year = {2013},
issn = {0010-4655},
doi = {https://doi.org/10.1016/j.cpc.2012.11.019},
url = {https://www.sciencedirect.com/science/article/pii/S0010465512003955},
author = {J.R. Johansson and P.D. Nation and Franco Nori}
}

@article{Peise2015Interaction,
	author = {Peise, J. and L{\"u}cke, B. and Pezz{\'e}, L. and Deuretzbacher, F. and Ertmer, W. and Arlt, J. and Smerzi, A. and Santos, L. and Klempt, C.},
	date = {2015/04/14},
	doi = {10.1038/ncomms7811},
	id = {Peise2015},
	isbn = {2041-1723},
	journal = {Nature Communications},
	number = {1},
	pages = {6811},
	title = {Interaction-free measurements by quantum Zeno stabilization of ultracold atoms},
	url = {https://doi.org/10.1038/ncomms7811},
	volume = {6},
	year = {2015}}

@article{Kruse2016Improvement,
  title = {Improvement of an Atomic Clock using Squeezed Vacuum},
  author = {Kruse, I. and Lange, K. and Peise, J. and L\"ucke, B. and Pezz\`e, L. and Arlt, J. and Ertmer, W. and Lisdat, C. and Santos, L. and Smerzi, A. and Klempt, C.},
  journal = {Phys. Rev. Lett.},
  volume = {117},
  issue = {14},
  pages = {143004},
  numpages = {5},
  year = {2016},
  month = {Sep},
  publisher = {American Physical Society},
  doi = {10.1103/PhysRevLett.117.143004},
  url = {https://link.aps.org/doi/10.1103/PhysRevLett.117.143004}
}

@article{Hamley2012Spin,
	author = {Hamley, C. D. and Gerving, C. S. and Hoang, T. M. and Bookjans, E. M. and Chapman, M. S.},
	date = {2012/04/01},
	id = {Hamley2012},
	isbn = {1745-2481},
	journal = {Nature Physics},
	number = {4},
	pages = {305--308},
	title = {Spin-nematic squeezed vacuum in a quantum gas},
	url = {https://doi.org/10.1038/nphys2245},
	volume = {8},
	year = {2012}
	}

@article{Linnemann2016Quantum,
  title = {Quantum-Enhanced Sensing Based on Time Reversal of Nonlinear Dynamics},
  author = {Linnemann, D. and Strobel, H. and Muessel, W. and Schulz, J. and Lewis-Swan, R. J. and Kheruntsyan, K. V. and Oberthaler, M. K.},
  journal = {Phys. Rev. Lett.},
  volume = {117},
  issue = {1},
  pages = {013001},
  numpages = {5},
  year = {2016},
  month = {Jun},
  publisher = {American Physical Society},
  url = {https://link.aps.org/doi/10.1103/PhysRevLett.117.013001}
}

@article{Sorelli2019Fast,
  title = {Fast and optimal generation of entanglement in bosonic Josephson junctions},
  author = {Sorelli, Giacomo and Gessner, Manuel and Smerzi, Augusto and Pezz\`e, Luca},
  journal = {Phys. Rev. A},
  volume = {99},
  issue = {2},
  pages = {022329},
  numpages = {11},
  year = {2019},
  month = {Feb},
  publisher = {American Physical Society},
  url = {https://link.aps.org/doi/10.1103/PhysRevA.99.022329}
}

@article{Helstrom1969Quantum,
	author = {Helstrom, Carl W. },
	date = {1969/06/01},
	doi = {10.1007/BF01007479},
	id = {Helstrom1969},
	isbn = {1572-9613},
	journal = {Journal of Statistical Physics},
	number = {2},
	pages = {231--252},
	title = {Quantum detection and estimation theory},
	url = {https://doi.org/10.1007/BF01007479},
	volume = {1},
	year = {1969}
}

@article{Braunstein1994Statistical,
  title = {Statistical distance and the geometry of quantum states},
  author = {Braunstein, Samuel L. and Caves, Carlton M.},
  journal = {Phys. Rev. Lett.},
  volume = {72},
  issue = {22},
  pages = {3439--3443},
  numpages = {0},
  year = {1994},
  month = {May},
  publisher = {American Physical Society},
  doi = {10.1103/PhysRevLett.72.3439},
  url = {https://link.aps.org/doi/10.1103/PhysRevLett.72.3439}
}

@article{Wineland1992Spin,
  title = {Spin squeezing and reduced quantum noise in spectroscopy},
  author = {Wineland, D. J. and Bollinger, J. J. and Itano, W. M. and Moore, F. L. and Heinzen, D. J.},
  journal = {Phys. Rev. A},
  volume = {46},
  issue = {11},
  pages = {R6797--R6800},
  numpages = {0},
  year = {1992},
  month = {Dec},
  publisher = {American Physical Society},
  doi = {10.1103/PhysRevA.46.R6797},
  url = {https://link.aps.org/doi/10.1103/PhysRevA.46.R6797}
}

@article{Pezze2009Entanglement,
  title = {Entanglement, Nonlinear Dynamics, and the Heisenberg Limit},
  author = {Pezz\'e, Luca and Smerzi, Augusto},
  journal = {Phys. Rev. Lett.},
  volume = {102},
  issue = {10},
  pages = {100401},
  numpages = {4},
  year = {2009},
  month = {Mar},
  publisher = {American Physical Society},
  doi = {10.1103/PhysRevLett.102.100401},
  url = {https://link.aps.org/doi/10.1103/PhysRevLett.102.100401}
}

@article{Hyllus2012Fisher,
  title = {Fisher information and multiparticle entanglement},
  author = {Hyllus, Philipp and Laskowski, Wies\l{}aw and Krischek, Roland and Schwemmer, Christian and Wieczorek, Witlef and Weinfurter, Harald and Pezz\'e, Luca and Smerzi, Augusto},
  journal = {Phys. Rev. A},
  volume = {85},
  issue = {2},
  pages = {022321},
  numpages = {10},
  year = {2012},
  month = {Feb},
  publisher = {American Physical Society},
  url = {https://link.aps.org/doi/10.1103/PhysRevA.85.022321}
}

@article{Iachello1996Algebraic,
    author = {Iachello, F. and Oss, S.},
    title = "{Algebraic approach to molecular spectra: Two‐dimensional problems}",
    journal = {The Journal of Chemical Physics},
    volume = {104},
    number = {18},
    pages = {6956-6963},
    year = {1996},
    month = {05},
    issn = {0021-9606},
    doi = {10.1063/1.471412}
}

@article{Cejnar2007Phase,
doi = {10.1088/1751-8113/40/4/001},
url = {https://dx.doi.org/10.1088/1751-8113/40/4/001},
year = {2007},
month = {jan},
publisher = {},
volume = {40},
number = {4},
pages = {581},
author = {Pavel Cejnar and Francesco Iachello},
title = {Phase structure of interacting boson models in arbitrary dimension},
journal = {Journal of Physics A: Mathematical and Theoretical}
}

@book{Iachello1995Algebraic,
    author = {Iachello, F and Levine, R D},
    title = "{Algebraic Theory of Molecules}",
    publisher = {Oxford University Press},
    year = {1995},
    month = {04},
    isbn = {9780195080919},
    doi = {10.1093/oso/9780195080919.001.0001},
    url = {https://doi.org/10.1093/oso/9780195080919.001.0001},
}

@article{Caprio2008Excited,
title = {Excited state quantum phase transitions in many-body systems},
journal = {Annals of Physics},
volume = {323},
number = {5},
pages = {1106-1135},
year = {2008},
issn = {0003-4916},
doi = {https://doi.org/10.1016/j.aop.2007.06.011},
url = {https://www.sciencedirect.com/science/article/pii/S0003491607001042},
author = {M.A. Caprio and P. Cejnar and F. Iachello}
}

@article{Cejnar2021Excited,
url = {https://doi.org/10.1088/1751-8121/abdfe8},
year = {2021},
month = {mar},
publisher = {IOP Publishing},
volume = {54},
number = {13},
pages = {133001},
author = {Cejnar, Pavel and Stránský, Pavel and Macek, Michal and Kloc, Michal},
title = {Excited-state quantum phase transitions},
journal = {Journal of Physics A: Mathematical and Theoretical}
}

@phdthesis{FeldmannThesis,
  author  = {Feldmann, Polina},
  title   = {Generalized quantum phase transitions for quantum-state engineering in spinor Bose-Einstein condensates},
  school  = {Gottfried Wilhelm Leibniz Universit\"{a}t},
  year    = {2021},
  doi = {https://doi.org/10.15488/10772}
}

@article{Feldmann2021Interferometric,
  title = {Interferometric Order Parameter for Excited-State Quantum Phase Transitions in Bose-Einstein Condensates},
  author = {Feldmann, Polina and Klempt, Carsten and Smerzi, Augusto and Santos, Luis and Gessner, Manuel},
  journal = {Phys. Rev. Lett.},
  volume = {126},
  issue = {23},
  pages = {230602},
  numpages = {6},
  year = {2021},
  month = {Jun},
  publisher = {American Physical Society},
  doi = {10.1103/PhysRevLett.126.230602},
  url = {https://link.aps.org/doi/10.1103/PhysRevLett.126.230602}
}

@book{Iachello1987The, 
place={Cambridge}, 
series={Cambridge Monographs on Mathematical Physics}, 
title={The Interacting Boson Model}, 
publisher={Cambridge University Press}, 
author={Iachello, F. and Arima, A.}, 
year={1987}, 
collection={Cambridge Monographs on Mathematical Physics}
}

@article{Kawaguchi2012Spinor,
title = {Spinor Bose–Einstein condensates},
journal = {Physics Reports},
volume = {520},
number = {5},
pages = {253-381},
year = {2012},
issn = {0370-1573},
doi = {https://doi.org/10.1016/j.physrep.2012.07.005},
url = {https://www.sciencedirect.com/science/article/pii/S0370157312002098},
author = {Yuki Kawaguchi and Masahito Ueda}
}

@article{Stamper2013Spinor,
  title = {Spinor Bose gases: Symmetries, magnetism, and quantum dynamics},
  author = {Stamper-Kurn, Dan M. and Ueda, Masahito},
  journal = {Rev. Mod. Phys.},
  volume = {85},
  issue = {3},
  pages = {1191--1244},
  numpages = {0},
  year = {2013},
  month = {Jul},
  publisher = {American Physical Society},
  doi = {10.1103/RevModPhys.85.1191},
  url = {https://link.aps.org/doi/10.1103/RevModPhys.85.1191}
}

@article{Anders2021Momentum,
  title = {Momentum Entanglement for Atom Interferometry},
  author = {Anders, F. and Idel, A. and Feldmann, P. and Bondarenko, D. and Loriani, S. and Lange, K. and Peise, J. and Gersemann, M. and Meyer-Hoppe, B. and Abend, S. and Gaaloul, N. and Schubert, C. and Schlippert, D. and Santos, L. and Rasel, E. and Klempt, C.},
  journal = {Phys. Rev. Lett.},
  volume = {127},
  issue = {14},
  pages = {140402},
  numpages = {7},
  year = {2021},
  month = {Sep},
  publisher = {American Physical Society},
  doi = {10.1103/PhysRevLett.127.140402},
  url = {https://link.aps.org/doi/10.1103/PhysRevLett.127.140402}
}

@article{Feldmann2018Interferometric,
  title = {Interferometric sensitivity and entanglement by scanning through quantum phase transitions in spinor Bose-Einstein condensates},
  author = {Feldmann, P. and Gessner, M. and Gabbrielli, M. and Klempt, C. and Santos, L. and Pezz\`e, L. and Smerzi, A.},
  journal = {Phys. Rev. A},
  volume = {97},
  issue = {3},
  pages = {032339},
  numpages = {10},
  year = {2018},
  month = {Mar},
  publisher = {American Physical Society},
  doi = {10.1103/PhysRevA.97.032339},
  url = {https://link.aps.org/doi/10.1103/PhysRevA.97.032339}
}

@article{Pezze2019Heralded,
  title = {Heralded Generation of Macroscopic Superposition States in a Spinor Bose-Einstein Condensate},
  author = {Pezz\`e, L. and Gessner, M. and Feldmann, P. and Klempt, C. and Santos, L. and Smerzi, A.},
  journal = {Phys. Rev. Lett.},
  volume = {123},
  issue = {26},
  pages = {260403},
  numpages = {7},
  year = {2019},
  month = {Dec},
  publisher = {American Physical Society},
  doi = {10.1103/PhysRevLett.123.260403},
  url = {https://link.aps.org/doi/10.1103/PhysRevLett.123.260403}
}

@article{Larese2013Signatures,
title = {Signatures of quantum phase transitions and excited state quantum phase transitions in the vibrational bending dynamics of triatomic molecules},
journal = {Journal of Molecular Structure},
volume = {1051},
pages = {310-327},
year = {2013},
issn = {0022-2860},
doi = {https://doi.org/10.1016/j.molstruc.2013.08.020},
url = {https://www.sciencedirect.com/science/article/pii/S0022286013006960},
author = {Danielle Larese and Francisco Pérez-Bernal and Francesco Iachello}
}

@article{Bookjans2011Quantum,
  title = {Quantum Phase Transition in an Antiferromagnetic Spinor Bose-Einstein Condensate},
  author = {Bookjans, E. M. and Vinit, A. and Raman, C.},
  journal = {Phys. Rev. Lett.},
  volume = {107},
  issue = {19},
  pages = {195306},
  numpages = {5},
  year = {2011},
  month = {Nov},
  publisher = {American Physical Society},
  doi = {10.1103/PhysRevLett.107.195306},
  url = {https://link.aps.org/doi/10.1103/PhysRevLett.107.195306}
}

@article{Zhang2013Generation,
  title = {Generation of Massive Entanglement through an Adiabatic Quantum Phase Transition in a Spinor Condensate},
  author = {Zhang, Z. and Duan, L.-M.},
  journal = {Phys. Rev. Lett.},
  volume = {111},
  issue = {18},
  pages = {180401},
  numpages = {5},
  year = {2013},
  month = {Oct},
  publisher = {American Physical Society},
  doi = {10.1103/PhysRevLett.111.180401},
  url = {https://link.aps.org/doi/10.1103/PhysRevLett.111.180401}
}

@article{Zhang2005Coherent,
  title = {Coherent spin mixing dynamics in a spin-1 atomic condensate},
  author = {Zhang, Wenxian and Zhou, D. L. and Chang, M.-S. and Chapman, M. S. and You, L.},
  journal = {Phys. Rev. A},
  volume = {72},
  issue = {1},
  pages = {013602},
  numpages = {5},
  year = {2005},
  month = {Jul},
  publisher = {American Physical Society},
  url = {https://link.aps.org/doi/10.1103/PhysRevA.72.013602}
}

@article{Zibold2010Classical,
  title = {Classical Bifurcation at the Transition from Rabi to Josephson Dynamics},
  author = {Zibold, Tilman and Nicklas, Eike and Gross, Christian and Oberthaler, Markus K.},
  journal = {Phys. Rev. Lett.},
  volume = {105},
  issue = {20},
  pages = {204101},
  numpages = {4},
  year = {2010},
  month = {Nov},
  publisher = {American Physical Society},
  doi = {10.1103/PhysRevLett.105.204101},
  url = {https://link.aps.org/doi/10.1103/PhysRevLett.105.204101}
}

@article{Zhao2014Dynamics,
  title = {Dynamics in spinor condensates tuned by a microwave dressing field},
  author = {Zhao, L. and Jiang, J. and Tang, T. and Webb, M. and Liu, Y.},
  journal = {Phys. Rev. A},
  volume = {89},
  issue = {2},
  pages = {023608},
  numpages = {5},
  year = {2014},
  month = {Feb},
  publisher = {American Physical Society},
  url = {https://link.aps.org/doi/10.1103/PhysRevA.89.023608}
}

@article{Mirkhalaf2021Criticality,
  title = {Criticality-enhanced quantum sensing in ferromagnetic Bose-Einstein condensates: Role of readout measurement and detection noise},
  author = {Mirkhalaf, Safoura S. and Benedicto Orenes, Daniel and Mitchell, Morgan W. and Witkowska, Emilia},
  journal = {Phys. Rev. A},
  volume = {103},
  issue = {2},
  pages = {023317},
  numpages = {15},
  year = {2021},
  month = {Feb},
  publisher = {American Physical Society},
  doi = {10.1103/PhysRevA.103.023317},
  url = {https://link.aps.org/doi/10.1103/PhysRevA.103.023317}
}

@article{PhysRevA.104.L031305,
  title = {Dynamical preparation of stripe states in spin-orbit-coupled gases},
  author = {Cabedo, J. and Claramunt, J. and Celi, A.},
  journal = {Phys. Rev. A},
  volume = {104},
  issue = {3},
  pages = {L031305},
  numpages = {7},
  year = {2021},
  month = {Sep},
  publisher = {American Physical Society},
  doi = {10.1103/PhysRevA.104.L031305},
  url = {https://link.aps.org/doi/10.1103/PhysRevA.104.L031305}
}

@article{PhysRevLett.124.043001,
  title = {Observation of Dynamical Quantum Phase Transitions with Correspondence in an Excited State Phase Diagram},
  author = {Tian, T. and Yang, H.-X. and Qiu, L.-Y. and Liang, H.-Y. and Yang, Y.-B. and Xu, Y. and Duan, L.-M.},
  journal = {Phys. Rev. Lett.},
  volume = {124},
  issue = {4},
  pages = {043001},
  numpages = {6},
  year = {2020},
  month = {Jan},
  publisher = {American Physical Society},
  doi = {10.1103/PhysRevLett.124.043001},
  url = {https://link.aps.org/doi/10.1103/PhysRevLett.124.043001}
}

@article{LiYouScience2017,
author = {Xin-Yu Luo  and Yi-Quan Zou  and Ling-Na Wu  and Qi Liu  and Ming-Fei Han  and Meng Khoon Tey  and Li You },
title = {Deterministic entanglement generation from driving through quantum phase transitions},
journal = {Science},
volume = {355},
number = {6325},
pages = {620-623},
year = {2017},
doi = {10.1126/science.aag1106}
}

@article{RevModPhys.90.035005,
  title = {Quantum metrology with nonclassical states of atomic ensembles},
  author = {Pezz\`e, Luca and Smerzi, Augusto and Oberthaler, Markus K. and Schmied, Roman and Treutlein, Philipp},
  journal = {Rev. Mod. Phys.},
  volume = {90},
  issue = {3},
  pages = {035005},
  numpages = {70},
  year = {2018},
  month = {Sep},
  publisher = {American Physical Society},
  doi = {10.1103/RevModPhys.90.035005},
  url = {https://link.aps.org/doi/10.1103/RevModPhys.90.035005}
}

@article{Bloch2012,
   author = {Immanuel Bloch and Jean Dalibard and Sylvain Nascimbène},
   doi = {10.1038/nphys2259},
   issn = {1745-2473},
   issue = {4},
   journal = {Nature Physics},
   month = {4},
   pages = {267-276},
   publisher = {Nature Publishing Group},
   title = {Quantum simulations with ultracold quantum gases},
   volume = {8},
   url = {http://www.nature.com/articles/nphys2259},
   year = {2012},
}

@article{Schafer2020,
	author = {Sch{\"a}fer, Florian and Fukuhara, Takeshi and Sugawa, Seiji and Takasu, Yosuke and Takahashi, Yoshiro},
	date = {2020/08/01},
	id = {Sch{\"a}fer2020},
	isbn = {2522-5820},
	journal = {Nature Reviews Physics},
	number = {8},
	pages = {411--425},
	title = {Tools for quantum simulation with ultracold atoms in optical lattices},
	url = {https://doi.org/10.1038/s42254-020-0195-3},
	volume = {2},
	year = {2020},
}

@article{Gross2017,
	author = {Christian Gross and Immanuel Bloch},
	doi = {10.1126/science.aal3837},
	journal = {Science},
	number = {6355},
	pages = {995-1001},
	title = {Quantum simulations with ultracold atoms in optical lattices},
	url = {https://www.science.org/doi/abs/10.1126/science.aal3837},
	volume = {357},
	year = {2017},
	}

@article{Jaksch2005,
   author = {Dieter Jaksch and P. Zoller},
   doi = {10.1016/J.AOP.2004.09.010},
   issn = {0003-4916},
   issue = {1},
   journal = {Annals of Physics},
   month = {1},
   pages = {52-79},
   publisher = {Academic Press},
   title = {The cold atom Hubbard toolbox},
   volume = {315},
   year = {2005},
}

@article{CiracNature2019,
	author = {Arg{\"u}ello-Luengo, Javier and Gonz{\'a}lez-Tudela, Alejandro and Shi, Tao and Zoller, Peter and Cirac, J. Ignacio},
	date = {2019/10/01},
	doi = {10.1038/s41586-019-1614-4},
	id = {Arg{\"u}ello-Luengo2019},
	isbn = {1476-4687},
	journal = {Nature},
	number = {7777},
	pages = {215--218},
	title = {Analogue quantum chemistry simulation},
	url = {https://doi.org/10.1038/s41586-019-1614-4},
	volume = {574},
	year = {2019}
}

@article{PhysRevX.8.031022,
  title = {Quantum Chemistry Calculations on a Trapped-Ion Quantum Simulator},
  author = {Hempel, Cornelius and Maier, Christine and Romero, Jonathan and McClean, Jarrod and Monz, Thomas and Shen, Heng and Jurcevic, Petar and Lanyon, Ben P. and Love, Peter and Babbush, Ryan and Aspuru-Guzik, Al\'an and Blatt, Rainer and Roos, Christian F.},
  journal = {Phys. Rev. X},
  volume = {8},
  issue = {3},
  pages = {031022},
  numpages = {22},
  year = {2018},
  month = {Jul},
  publisher = {American Physical Society},
  doi = {10.1103/PhysRevX.8.031022},
  url = {https://link.aps.org/doi/10.1103/PhysRevX.8.031022}
}

@article{PhysRevX.6.031007,
  title = {Scalable Quantum Simulation of Molecular Energies},
  author = {O'Malley, P. J. J. and Babbush, R. and Kivlichan, I. D. and Romero, J. and McClean, J. R. and Barends, R. and Kelly, J. and Roushan, P. and Tranter, A. and Ding, N. and Campbell, B. and Chen, Y. and Chen, Z. and Chiaro, B. and Dunsworth, A. and Fowler, A. G. and Jeffrey, E. and Lucero, E. and Megrant, A. and Mutus, J. Y. and Neeley, M. and Neill, C. and Quintana, C. and Sank, D. and Vainsencher, A. and Wenner, J. and White, T. C. and Coveney, P. V. and Love, P. J. and Neven, H. and Aspuru-Guzik, A. and Martinis, J. M.},
  journal = {Phys. Rev. X},
  volume = {6},
  issue = {3},
  pages = {031007},
  numpages = {13},
  year = {2016},
  month = {Jul},
  publisher = {American Physical Society},
  doi = {10.1103/PhysRevX.6.031007},
  url = {https://link.aps.org/doi/10.1103/PhysRevX.6.031007}
}

@article{Daley2022,
   author = {Andrew J. Daley and Immanuel Bloch and Christian Kokail and Stuart Flannigan and Natalie Pearson and Matthias Troyer and Peter Zoller},
   doi = {10.1038/s41586-022-04940-6},
   issn = {1476-4687},
   issue = {7920},
   journal = {Nature},
   month = {7},
   pages = {667-676},
   pmid = {35896643},
   publisher = {Nature Publishing Group},
   title = {Practical quantum advantage in quantum simulation},
   volume = {607},
   url = {https://www.nature.com/articles/s41586-022-04940-6},
   year = {2022},
}

@article{Schneider2012,
   author = {Ch Schneider and Diego Porras and Tobias Schaetz},
   doi = {10.1088/0034-4885/75/2/024401},
   issn = {00344885},
   issue = {2},
   journal = {Reports on Progress in Physics},
   month = {2},
   pages = {024401},
   publisher = {IOP Publishing},
   title = {Experimental quantum simulations of many-body physics with trapped ions},
   volume = {75},
   url = {http://stacks.iop.org/0034-4885/75/i=2/a=024401?key=crossref.edb9862d303e788fb963b3583ea41f33},
   year = {2012},
}

@article{Schlawin2020,
  title = {Continuously Parametrized Quantum Simulation of Molecular Electron-Transfer Reactions},
  author = {Schlawin, Frank and Gessner, Manuel and Buchleitner, Andreas and Sch\"atz, Tobias and Skourtis, Spiros S.},
  journal = {PRX Quantum},
  volume = {2},
  issue = {1},
  pages = {010314},
  numpages = {12},
  year = {2021},
  month = {Jan},
  publisher = {American Physical Society},
  doi = {10.1103/PRXQuantum.2.010314},
  url = {https://link.aps.org/doi/10.1103/PRXQuantum.2.010314}
}

@article{Devoret2024,
  title = {Asymmetry Control in a Parametric Oscillator for the Quantum Simulation of Chemical Activation},
  author = {de Albornoz, Alejandro Cros Carrillo and Corti\~nas, Rodrigo G. and Sch\"afer, Max and Frattini, Nicholas E. and Allen, Brandon and Cabral, Delmar G.A. and Videla, Pablo E. and Khazaei, Pouya and Geva, Eitan and Batista, Victor S. and Devoret, Michel H.},
  journal = {PRX Quantum},
  volume = {7},
  issue = {2},
  pages = {020309},
  numpages = {18},
  year = {2026},
  month = {Apr},
  publisher = {American Physical Society},
  doi = {10.1103/71yp-fqns},
  url = {https://link.aps.org/doi/10.1103/71yp-fqns}
}

@article{So2024,
	author = {Visal So and Midhuna Duraisamy Suganthi and Abhishek Menon and Mingjian Zhu and Roman Zhuravel and Han Pu and Peter G. Wolynes and Jos{\'e} N. Onuchic and Guido Pagano},
	doi = {10.1126/sciadv.ads8011},
	journal = {Science Advances},
	number = {51},
	pages = {eads8011},
	title = {Trapped-ion quantum simulation of electron transfer models with tunable dissipation},
	url = {https://www.science.org/doi/abs/10.1126/sciadv.ads8011},
	volume = {10},
	year = {2024}
    }

@article{Cabedo2021,
  title = {Excited-state quantum phase transitions in spin-orbit-coupled Bose gases},
  author = {Cabedo, J. and Celi, A.},
  journal = {Phys. Rev. Res.},
  volume = {3},
  issue = {4},
  pages = {043215},
  numpages = {12},
  year = {2021},
  month = {Dec},
  publisher = {American Physical Society},
  doi = {10.1103/PhysRevResearch.3.043215},
  url = {https://link.aps.org/doi/10.1103/PhysRevResearch.3.043215}
}

@article{Chisholm2024,
	author = {C. S. Chisholm and S. Hirthe and V. B. Makhalov and R. Ramos and R. Vatr{\'e} and J. Cabedo and A. Celi and L. Tarruell},
	doi = {10.1126/science.adv1209},
	journal = {Science},
	number = {6784},
	pages = {480-484},
	title = {Probing supersolidity through excitations in a spin-orbit--coupled Bose-Einstein condensate},
	volume = {391},
	year = {2026}
}

@article{PhysRevA.6.2211,
  title = {Atomic Coherent States in Quantum Optics},
  author = {Arecchi, F. T. and Courtens, Eric and Gilmore, Robert and Thomas, Harry},
  journal = {Phys. Rev. A},
  volume = {6},
  issue = {6},
  pages = {2211--2237},
  numpages = {0},
  year = {1972},
  month = {Dec},
  publisher = {American Physical Society},
  doi = {10.1103/PhysRevA.6.2211},
  url = {https://link.aps.org/doi/10.1103/PhysRevA.6.2211}
}

@article{PhysRev.131.2766,
   author = {Roy J. Glauber},
   doi = {10.1103/PhysRev.131.2766},
   isbn = {0031-899X},
   issn = {0031899X},
   issue = {6},
   journal = {Phys. Rev.},
   pages = {2766-2788},
   pmid = {15648836249804302126},
   publisher = {American Physical Society},
   title = {Coherent and incoherent states of the radiation field},
   volume = {131},
   url = {http://link.aps.org/doi/10.1103/PhysRev.131.2766},
   year = {1963},
}

@article{Radcliffe1971,
   author = {J. M. Radcliffe},
   doi = {10.1088/0305-4470/4/3/009},
   isbn = {0022-3689},
   issn = {03054470},
   issue = {3},
   journal = {Journal of Physics A},
   pages = {313-323},
   title = {Some properties of coherent spin states},
   volume = {4},
   url = {http://stacks.iop.org/0022-3689/4/i=3/a=009},
   year = {1971},
}

@book{MandelWolf,
   author = {L. Mandel and Emil Wolf},
   city = {Cambridge, UK},
   isbn = {0521417112},
   publisher = {Cambridge University Press},
   title = {Optical Coherence and Quantum Optics},
   year = {1995},
}

@article{PhysRevA.50.67,
  title = {Squeezed atomic states and projection noise in spectroscopy},
  author = {Wineland, D. J. and Bollinger, J. J. and Itano, W. M. and Heinzen, D. J.},
  journal = {Phys. Rev. A},
  volume = {50},
  issue = {1},
  pages = {67--88},
  numpages = {0},
  year = {1994},
  month = {Jul},
  publisher = {American Physical Society},
  doi = {10.1103/PhysRevA.50.67},
  url = {https://link.aps.org/doi/10.1103/PhysRevA.50.67}
}

@article{Gessner2019,
  title = {Metrological Nonlinear Squeezing Parameter},
  author = {Gessner, Manuel and Smerzi, Augusto and Pezz\`e, Luca},
  journal = {Phys. Rev. Lett.},
  volume = {122},
  issue = {9},
  pages = {090503},
  numpages = {7},
  year = {2019},
  month = {Mar},
  publisher = {American Physical Society},
  doi = {10.1103/PhysRevLett.122.090503},
  url = {https://link.aps.org/doi/10.1103/PhysRevLett.122.090503}
}

@article{PhysRevA.85.022321,
  title = {Fisher information and multiparticle entanglement},
  author = {Hyllus, Philipp and Laskowski, Wies\l{}aw and Krischek, Roland and Schwemmer, Christian and Wieczorek, Witlef and Weinfurter, Harald and Pezz\'e, Luca and Smerzi, Augusto},
  journal = {Phys. Rev. A},
  volume = {85},
  issue = {2},
  pages = {022321},
  numpages = {10},
  year = {2012},
  month = {Feb},
  publisher = {American Physical Society},
  url = {https://link.aps.org/doi/10.1103/PhysRevA.85.022321}
}

@article{Ren2021,
  title = {Metrological Detection of Multipartite Entanglement from Young Diagrams},
  author = {Ren, Zhihong and Li, Weidong and Smerzi, Augusto and Gessner, Manuel},
  journal = {Phys. Rev. Lett.},
  volume = {126},
  issue = {8},
  pages = {080502},
  numpages = {6},
  year = {2021},
  month = {Feb},
  publisher = {American Physical Society},
  url = {https://link.aps.org/doi/10.1103/PhysRevLett.126.080502}
}

@article{Riedel2010Atom,
	author = {Riedel, Max F. and B{\"o}hi, Pascal and Li, Yun and H{\"a}nsch, Theodor W. and Sinatra, Alice and Treutlein, Philipp},
	date = {2010/04/01},
	id = {Riedel2010},
	isbn = {1476-4687},
	journal = {Nature},
	number = {7292},
	pages = {1170--1173},
	title = {Atom-chip-based generation of entanglement for quantum metrology},
	url = {https://doi.org/10.1038/nature08988},
	volume = {464},
	year = {2010}
}

@article{McConnell2015Entanglement,
	author = {McConnell, Robert and Zhang, Hao and Hu, Jiazhong and {\'C}uk, Senka and Vuleti{\'c}, Vladan},
	date = {2015/03/01},
	id = {McConnell2015},
	isbn = {1476-4687},
	journal = {Nature},
	number = {7544},
	pages = {439--442},
	title = {Entanglement with negative Wigner function of almost 3,000 atoms heralded by one photon},
	url = {https://doi.org/10.1038/nature14293},
	volume = {519},
	year = {2015}
}

@article{Meyer2023Excited,
  title = {Excited-State Phase Diagram of a Ferromagnetic Quantum Gas},
  author = {Meyer-Hoppe, B. and Anders, F. and Feldmann, P. and Santos, L. and Klempt, C.},
  journal = {Phys. Rev. Lett.},
  volume = {131},
  issue = {24},
  pages = {243402},
  numpages = {6},
  year = {2023},
  month = {Dec},
  publisher = {American Physical Society},
  doi = {10.1103/PhysRevLett.131.243402},
  url = {https://link.aps.org/doi/10.1103/PhysRevLett.131.243402}
}

@Article{Raggio1989Quantum,
author={Raggio, G. A.
and Werner, R. F.},
title={Quantum statistical mechanics of general mean field systems},
journal={Helvetica Physica Acta},
year={1989},
publisher={Birkh{\"a}user},
volume={62},
number={8},
pages={980},
issn={0018-0238},
doi={10.5169/seals-116175},
url={https://doi.org/10.5169/seals-116175}
}

@article{Duffield1992Mean,
author = {Duffield, N. G. and Werner, R. F.},
title = {Mean-Field dynamical semigroups on C*-algebras},
journal = {Reviews in Mathematical Physics},
volume = {04},
number = {03},
pages = {383-424},
year = {1992},
doi = {10.1142/S0129055X92000108},
URL = {https://doi.org/10.1142/S0129055X92000108}
}

@inbook{Duffield1992Classical,
title = "Classical Hamiltonian dynamics for quantum Hamiltonian mean-field limits",
author = "Duffield, N. G. and Werner, R. F.",
year = "1992",
isbn = "981-02-1015-9",
pages = "115--129",
editor = "A Truman and Davies, {I. M.}",
booktitle = "Stochastics and quantum mechanics (Swansea, 1990)",
publisher = "World Sci. Publishing",
}

@article{Strobel2014Fisher,
author = {Helmut Strobel  and Wolfgang Muessel  and Daniel Linnemann  and Tilman Zibold  and David B. Hume  and Luca Pezzè  and Augusto Smerzi  and Markus K. Oberthaler },
title = {Fisher information and entanglement of non-Gaussian spin states},
journal = {Science},
volume = {345},
number = {6195},
pages = {424-427},
year = {2014},
doi = {10.1126/science.1250147}
}

@misc{Usui2025Code,
  title        = "code for Spinor Bose-Einstein condensate as an analog simulator of molecular bending vibrations",
  url = "https://doi.org/10.5281/zenodo.23017382",
  author       = "Usui, Ayaka and Niezgoda, Artur",
  year         = "2026"
}

@article{Khalouf2021Excited,
title = {Excited state quantum phase transitions in the bending spectra of molecules},
journal = {Journal of Quantitative Spectroscopy and Radiative Transfer},
volume = {261},
pages = {107436},
year = {2021},
issn = {0022-4073},
doi = {https://doi.org/10.1016/j.jqsrt.2020.107436},
url = {https://www.sciencedirect.com/science/article/pii/S002240732030964X},
author = {Jamil Khalouf-Rivera and Francisco Pérez-Bernal and Miguel Carvajal}
}

@article{Khalouf2019Calculation,
	author = {Khalouf-Rivera, Jamil and Carvajal, Miguel and Santos, Lea F. and P{\'e}rez-Bernal, Francisco},
	date = {2019/11/07},
	doi = {10.1021/acs.jpca.9b07338},
	isbn = {1089-5639},
	journal = {The Journal of Physical Chemistry A},
	journal1 = {The Journal of Physical Chemistry A},
	journal2 = {J. Phys. Chem. A},
	month = {11},
	number = {44},
	pages = {9544--9551},
	publisher = {American Chemical Society},
	title = {Calculation of Transition State Energies in the HCN--HNC Isomerization with an Algebraic Model},
	type = {doi: 10.1021/acs.jpca.9b07338},
	url = {https://doi.org/10.1021/acs.jpca.9b07338},
	volume = {123},
	year = {2019},
	year1 = {2019}}

@article{Khalouf2022Anharmonicity,
  title = {Anharmonicity-induced excited-state quantum phase transition in the symmetric phase of the two-dimensional limit of the vibron model},
  author = {Khalouf-Rivera, Jamil and P\'erez-Bernal, Francisco and Carvajal, Miguel},
  journal = {Phys. Rev. A},
  volume = {105},
  issue = {3},
  pages = {032215},
  numpages = {13},
  year = {2022},
  month = {Mar},
  publisher = {American Physical Society},
  doi = {10.1103/PhysRevA.105.032215},
  url = {https://link.aps.org/doi/10.1103/PhysRevA.105.032215}
}

@Article{Khalouf2022Quantum,
	title={{Quantum fidelity susceptibility in excited state quantum phase transitions: application to the bending spectra of nonrigid molecules}},
	author={Jamil Khalouf-Rivera and Miguel Carvajal and Francisco Pérez-Bernal},
	journal={SciPost Phys.},
	volume={12},
	pages={002},
	year={2022},
	publisher={SciPost},
	doi={10.21468/SciPostPhys.12.1.002},
	url={https://scipost.org/10.21468/SciPostPhys.12.1.002},
}

@article{Khalouf2024Degeneracy,
  title = {Degeneracy in excited-state quantum phase transitions of two-level bosonic models and its influence on system dynamics},
  author = {Khalouf-Rivera, Jamil and Wang, Qian and Santos, Lea F. and Garc\'{\i}a-Ramos, Jos\'e-Enrique and Carvajal, Miguel and P\'erez-Bernal, Francisco},
  journal = {Phys. Rev. A},
  volume = {109},
  issue = {6},
  pages = {062219},
  numpages = {14},
  year = {2024},
  month = {Jun},
  publisher = {American Physical Society},
  doi = {10.1103/PhysRevA.109.062219},
  url = {https://link.aps.org/doi/10.1103/PhysRevA.109.062219}
}

@book{Iachello2015Lie,
    series={Lecture Notes in Physics}, 
    title={Lie Algebras and Applications}, 
    publisher={Springer Berlin, Heidelberg}, 
    author={Francesco Iachello}, 
    year={2015}, 
    collection={Lecture Notes in Physics},
    doi = {10.1007/978-3-662-44494-8},
    url = {https://doi.org/10.1007/978-3-662-44494-8}
}

@article{Larese2011study,
title = {A study of quantum phase transitions and quantum monodromy in the bending motion of non-rigid molecules},
journal = {Journal of Molecular Structure},
volume = {1006},
number = {1},
pages = {611-628},
year = {2011},
issn = {0022-2860},
doi = {https://doi.org/10.1016/j.molstruc.2011.10.016},
url = {https://www.sciencedirect.com/science/article/pii/S0022286011008088},
author = {Danielle Larese and Francesco Iachello}
}

@article{Cejnar2008Impact,
  title = {Impact of quantum phase transitions on excited-level dynamics},
  author = {Cejnar, Pavel and Str\'ansk\'y, Pavel},
  journal = {Phys. Rev. E},
  volume = {78},
  issue = {3},
  pages = {031130},
  numpages = {5},
  year = {2008},
  month = {Sep},
  publisher = {American Physical Society},
  doi = {10.1103/PhysRevE.78.031130},
  url = {https://link.aps.org/doi/10.1103/PhysRevE.78.031130}
}

@article{Chavez2023Spectral,
	author = {Ch{\'a}vez-Carlos, Jorge and Lezama, Tal{\'\i}a L. M. and Corti{\~n}as, Rodrigo G. and Venkatraman, Jayameenakshi and Devoret, Michel H. and Batista, Victor S. and P{\'e}rez-Bernal, Francisco and Santos, Lea F.},
	date = {2023/07/28},
	doi = {10.1038/s41534-023-00745-1},
	id = {Ch{\'a}vez-Carlos2023},
	isbn = {2056-6387},
	journal = {npj Quantum Information},
	number = {1},
	pages = {76},
	title = {Spectral kissing and its dynamical consequences in the squeeze-driven Kerr oscillator},
	url = {https://doi.org/10.1038/s41534-023-00745-1},
	volume = {9},
	year = {2023}}

@incollection{Pezze2014Quantum,
  title={Quantum theory of phase estimation},
  author={Pezz{\`e}, L and Smerzi, A},
  series={Proceedings of the International School of Physics "Enrico Fermi"},
  booktitle={Atom Interferometry},
  pages={691--741},
  year={2014},
  doi={10.3254/978-1-61499-448-0-691},
  publisher={IOS Press}
}

% \clearpage
\appendix
\section{Algebraic models in two dimensions}\label{app:algebraic}

The generators of the algebra $U(3)$ are the bilinear products of these creation and
annihilation operators and can be written as: 
\begin{align}\label{eq:3moleculealgebra}
    \hat{n}
    &=
    \hat{\tau}_+^{\dagger}\hat{\tau}_+ + \hat{\tau}_-^{\dagger}\hat{\tau}_-
    ,
    \\ \nonumber
    \hat{l}
    &=
    \hat{\tau}_+^{\dagger}\hat{\tau}_+ - \hat{\tau}_-^{\dagger}\hat{\tau}_-
    ,
    \\ \nonumber
    \hat{Q}_+
    &=
    \sqrt{2}\hat{\tau}_+^{\dagger}\hat{\tau}_-
    ,
    \\ \nonumber
    \hat{Q}_-
    &=
    \sqrt{2}\hat{\tau}_-^{\dagger}\hat{\tau}_+
    ,
    \\ \nonumber
    \hat{n}_s
    &=
    \hat{\sigma}^{\dagger}\hat{\sigma}
    ,
    \\ \nonumber
    \hat{D}_+
    &=
    \sqrt{2}
    \left(
    \hat{\tau}_+^{\dagger}\hat{\sigma} - \hat{\sigma}^{\dagger}\hat{\tau}_-
    \right)
    ,
    \\ \nonumber
    \hat{D}_-
    &=
    \sqrt{2}
    \left(
    -\hat{\tau}_-^{\dagger}\hat{\sigma} + \hat{\sigma}^{\dagger}\hat{\tau}_+
    \right)
    ,
    \\ \nonumber
    \hat{R}_+
    &=
    \sqrt{2}\left(
    \hat{\tau}_+^{\dagger}\hat{\sigma} + \hat{\sigma}^{\dagger}\hat{\tau}_-
    \right)
    ,
    \\ \nonumber
    \hat{R}_-
    &=
    \sqrt{2}\left(
    \hat{\tau}_-^{\dagger}\hat{\sigma} + \hat{\sigma}^{\dagger}\hat{\tau}_+
    \right)
    .
\end{align}
We can distinguish two possible subalgebra chains within the Lie algebra, starting from $U(3)$ and ending in $SO(2)$, that conserve two-dimensional angular momentum:
\begin{subequations} \label{eq:subslgebras}
\begin{align}
    (I)&\hspace{1 cm} 
    U(3) \supset U(2) \supset SO(2)
    \label{eq:chainI}
    \\
    (II)&\hspace{1 cm} 
    U(3) \supset SO(3) \supset SO(2)
    \label{eq:chainII}
    .
\end{align}
\end{subequations}
The subalgebra $U(2)$ consists of the four generators $\{\hat{n},\hat{l},\hat{Q}_+,\hat{Q}_-\}$, while $O(3)$ is generated by three elements $\{\hat{D}_+,\hat{D}_-,\hat{l}\}$, which have similar commutation relations to angular momentum, $[\hat{l},\hat{D}_{\pm}]=\pm\hat{D}_{\pm}$ and $[\hat{D}_+,\hat{D}_-]=2\hat{l}$. 
Lastly, the algebra $O(2)$ consists of $\{\hat{l}\}$ which is the angular momentum in two dimensions, in $z$ direction on the $x,y$ plane. Note that the generators of $O(n)$ coincide with those of $SO(n)$ which is why we do not make a distinction in the notation here.

To construct a Hamiltonian that reflects either one of the two chains, the starting point of the algebraic approach is the Casimir operators for each of the subalgebras, i.e., those operators that commute with all generators. These operators may exist as linear, quadratic or higher-order functions of the generators. A generic algebraic Hamiltonian then consists of a linear combination of all Casimir operators. By taking into account up to quadratic terms, we obtain the following form, which encompasses both chains (I) and (II):
\begin{align} 
    \hat{H}
    &=
    E_0 
    + \epsilon \hat{C}_1[U(2)]
    + \alpha \hat{C}_2[U(2)]
    \nonumber\\
    &\quad 
    + \beta \hat{C}_2[O(2)]
    + A \hat{C}_2[O(3)]
    \nonumber\\
    &=
    E_0 
    + \epsilon\hat{n}
    + \alpha \hat{n}\left(\hat{n}+1\right)
    + \beta \hat{l}^2
    + A \hat{W}^2
    ,
\end{align}
where $\hat{C}_{1,2}$ is the linear or quadratic invariant Casimir operator for a subalgebra of $U(3)$ 
\begin{align}
    \hat{C}_1[U(2)] &= \hat{n}  \hspace{1cm } \hat{C}_2[U(2)] = \hat{n}(\hat{n}+1)\\ \nonumber
    \hat{C}_2[O(2)] &= \hat{l}^2  \hspace{1cm } \hat{C}_2[O(3)] = \hat{W}^2,
\end{align}
and 
\begin{align}
    \hat{W}^2
    &=
    \frac{1}{2}\left(
    \hat{D}_+\hat{D}_- + \hat{D}_-\hat{D}_+
    \right)
    +
    \hat{l}^2
    .
\end{align}
Note that there is no linear Casimir operator for $O(2)$ and $U(3)$ in the two-dimensional case. 

\subsection{Chain (I)} \label{app:algebraic_chainI}

In the subalgebra chain (I) in~\eqref{eq:chainI}, the basis is determined by the quantum numbers as
\begin{subequations}
\begin{align}
    U(3) &\to N \\
    U(2) &\to n \\
    O(2) &\to l
    .
\end{align}
\end{subequations}
These numbers characterise the states in this chain and have physical meaning. Here, $N$ is the total number of bound states, $n$ is the vibrational quantum number, and $l$ is the value of the angular momentum along the $z$-axis.  These quantum numbers are bounded by the quantum number of the upper group, 
i.e. $n=0,1,\ldots,N$ and $l=-n,-n+2,\ldots,n-2,n$.
The Hamiltonian with the dynamical symmetry is given by
\begin{align} \label{eq:HchainI}
    \hat{H}^{(I)}
    &=
    E_0 
    + \epsilon\hat{n}
    + \alpha \hat{n}\left(\hat{n}+1\right)
    + \beta \hat{l}^2
    ,
\end{align}
with eigenvalues
\begin{align}
    E^{(I)}
    &=
    E_0 
    + \epsilon n
    + \alpha n \left(n+1\right)
    + \beta l^2
    .
\end{align}
This Hamiltonian describes oscillations in two dimensions, particularly vibrations of linear polyatomic molecules. The same spectrum is obtained by solving the Schr\"{o}dinger equation with the P\"{o}schl-Teller potential
\begin{align}
    V(r)
    &=
    \frac{-V_0}{\cosh^2 r/a}
    ,
\end{align}
where $x$ and $y$ denote the transverse coordinates and $z$ is the molecular axis (see Fig.~\ref{fig:coordinates}), $r\equiv\sqrt{x^2+y^2}$, and $a$ is a bond length which is normalised into 1 in this work. 
This potential takes the minimum at $r=0$ implying that a linear configuration minimises the energy.

\subsection{Chain (II)} \label{app:algebraic_chainII}

In chain (II) in ~\eqref{eq:chainII}, we identify the following quantum numbers:
\begin{subequations}
\begin{align}
    U(3)& \to N \\
    O(3)& \to \omega \\
    O(2)& \to l
    .
\end{align}
\end{subequations}
The quantum numbers are bounded in a similar way to chain (I): $\omega=0,2\ldots,N$ or $\omega=1,3\ldots,N$ for even or odd $N$ and $l=-\omega,-\omega+1,\ldots,\omega-1,\omega$.
For the sake of convenience, we introduce $v=(N-\omega)/2$ instead of $\omega$, and $v$ is the vibrational quantum number of chain (II) and can be interpreted as the number of excitations of the displaced oscillator.
The quantum numbers are given by $v=0,1,\ldots,(N-1)/2$ or $N/2$ for odd or even $N$ and $l=-(N-2v),-(N-2v)+1,\ldots,N-2v-1,N-2v$.
The Hamiltonian with the corresponding dynamical symmetry is given by
\begin{align} \label{eq:HchainII}
    \hat{H}^{(II)}
    &=
    E_0 
    + \beta \hat{l}^2
    + A \hat{W}^2
\end{align}
with eigenvalues
\begin{align}
    E^{(II)}
    &=
    E_0 
    + \beta l^2
    + A \omega\left(\omega+1\right)
    \\
    &=
    E_0
    + AN(N+1)
    \nonumber\\
    &\quad\quad
    - 4A\left[
    \left(N+\frac{1}{2}\right)v-v^2
    \right]
    + \beta l^2\nonumber
    .
\end{align}
This Hamiltonian describes rotation and vibration of an atom of a bent molecule, and the energy spectrum corresponds to that of the two-dimensional Morse potential, 
\begin{align}
    V(r)
    &=
    V_0\left(
    \mathrm{e}^{-2r/a} - 2\mathrm{e}^{-r/a}
    \right)
    .
\end{align}
This potential has its minimum at $r=a$, which means that the atomic configuration is stable when the molecule is bent. 

\subsection{Intermediate regime chain I and chain II}

The intermediate regime between these two chains can be studied by diagonalising the general Hamiltonian~\eqref{eq:H2d}. 
Nevertheless, the phase transition between the two chains is captured by only two terms. 
The Casimir operator for $O(2)$ is shared by both chains and thus does not contribute to the phase transition. The linear and quadratic Casimir operators for $U(2)$ can be diagonalised in the same basis, and therefore we can limit our attention to the linear term. 
Thus, the essential part for the phase transition is given by 
\begin{align} 
    \hat{H}
    &=
    (1-\gamma)
    \hat{n}
    -
    \frac{\gamma}{N-1}
    \hat{W}^2
    ,
\end{align}
where $\gamma$ is a control parameter and the factor $1/(N-1)$ is introduced considering that $\hat{n}$ is a one-body operator that scales with $N$ and $\hat{W}^2$ is a two-body operator, scaling with $N(N-1)$.

\section{Spin-1 algebras}\label{app:spin1algebras}

We summarise the operators of spin-1 algebras used in the text. The $SU(3)$ group is spanned by 8 generators $\{\hat{J}_x, \hat{J}_y, \hat{J}_z, \hat{Q}_{xy}, \hat{Q}_{yz}, \hat{Q}_{zx}, \hat{Y}, \hat{D}_{xy}\}$, defined as
\begin{align}
    \hat{J}_x
    &\equiv
    \frac{1}{\sqrt{2}}\left(
    \hat{\sigma}^{\dagger} \hat{\tau}_+
    +
    \hat{\sigma}^{\dagger} \hat{\tau}_-
    +
    \hat{\tau}_+^{\dagger} \hat{\sigma}
    +
    \hat{\tau}_-^{\dagger} \hat{\sigma}
    \right),
    \\
    \hat{J}_y
    &\equiv
    \frac{i}{\sqrt{2}}\left(
    -\hat{\sigma}^{\dagger} \hat{\tau}_+
    +
    \hat{\sigma}^{\dagger} \hat{\tau}_-
    +
    \hat{\tau}_+^{\dagger} \hat{\sigma}
    -
    \hat{\tau}_-^{\dagger} \hat{\sigma}
    \right),
    \\
    \hat{J}_z
    &\equiv
    \hat{\tau}_+^{\dagger} \hat{\tau}_+
    -
    \hat{\tau}_-^{\dagger} \hat{\tau}_-,
\end{align}
\begin{align}
    \hat{Q}_{xy}
    &\equiv
    i\left(
    \hat{\tau}_+^{\dagger}\hat{\tau}_- 
    -
    \hat{\tau}_-^{\dagger}\hat{\tau}_+
    \right),
    \\
    \hat{Q}_{yz}
    &\equiv
    \frac{i}{\sqrt{2}}
    \left(
    -\hat{\sigma}^{\dagger}\hat{\tau}_-
    +\hat{\tau}_-^{\dagger}\hat{\sigma}
    +\hat{\tau}_+^{\dagger}\hat{\sigma}
    -\hat{\sigma}^{\dagger}\hat{\tau}_+
    \right),
    \\
    \hat{Q}_{zx}
    &\equiv
    \frac{1}{\sqrt{2}}
    \left(
    -\hat{\sigma}^{\dagger}\hat{\tau}_-
    -\hat{\tau}_-^{\dagger}\hat{\sigma}
    +\hat{\tau}_+^{\dagger}\hat{\sigma}
    +\hat{\sigma}^{\dagger}\hat{\tau}_+
    \right),
\end{align}
\begin{align}
    \hat{Y}
    &\equiv
    \frac{1}{\sqrt{3}}\left(
    \hat{\tau}_+^{\dagger}\hat{\tau}_+
    +
    \hat{\tau}_-^{\dagger}\hat{\tau}_-
    -2\hat{\sigma}^{\dagger}\hat{\sigma}
    \right),
    \\
    \hat{D}_{xy}
    &\equiv
    \hat{\tau}_+^{\dagger}\hat{\tau}_- 
    +
    \hat{\tau}_-^{\dagger}\hat{\tau}_+
    .
\end{align}
The mapping between the $SU(3)$ operators and those given in Eq.~\eqref{eq:3moleculealgebra} is straightforward:
\begin{align} \label{eq:spinor_2dvm}
    &\hat{N}_0 = \hat{n}_s, 
    \\ \nonumber
    &\hat{N} = \hat{n} + \hat{n}_s,
    \\ \nonumber
    &\hat{J}_x = \frac{1}{2}(\hat{R}_+ + \hat{R}_- ),
    \\ \nonumber
    &\hat{J}_y = \frac{i}{2}(\hat{R}_- - \hat{R}_+),
    \\ \nonumber
    &\hat{J}_z = \hat{l},
    \\ \nonumber
    &\hat{Q}_{zx} = \frac{1}{2}(\hat{D}_+ + \hat{D}_-),
    \\ \nonumber
    &\hat{Q}_{yz} = 
    \frac{i}{2}(\hat{D}_+ - \hat{D}_-),
    \\ \nonumber
    &\hat{Q}_{xy} = \frac{i}{\sqrt{2}}(\hat{Q}_{-} - \hat{Q}_+),
    \\ \nonumber
    &\hat{D}_{xy} = \frac{1}{\sqrt{2}}(\hat{Q}_{+} + \hat{Q}_-), 
    \\ \nonumber
    &\hat{Y} = \frac{1}{\sqrt{3}} (\hat{n} - 2\hat{n}_s)
    .
\end{align}
The $SU(3)$ algebra contains $SU(2)$ as a subalgebra. The paradigmatic constructions of $SU(2)$ subalgebras are based on symmetric and anti-symmetric subspaces, with bosonic annihilation operators defined as
\begin{subequations}
\begin{align}
    \hat{\tau}_s
    &\equiv
    \frac{\hat{\tau}_++\hat{\tau}_-}{\sqrt{2}}
    , \label{eq:gs}
    \\
    \hat{\tau}_a
    &\equiv
    \frac{\hat{\tau}_+-\hat{\tau}_-}{\sqrt{2}}
    . \label{eq:ga}
\end{align}
\end{subequations}
Here we use the subspaces of mode $x$ and of mode $y$, respectively. As a reminder, their annihilation operators are given by
\begin{subequations} \label{eq:gp_gm}
\begin{align}
    \hat{\tau}_x
    &\equiv
    -\frac{\hat{\tau}_+-\hat{\tau}_-}{\sqrt{2}}
    , \label{eq:gp}
    \\
    \hat{\tau}_y
    &\equiv
    -i\frac{\hat{\tau}_++\hat{\tau}_-}{\sqrt{2}}
    . \label{eq:gm}
\end{align}
\end{subequations}
These operators are similar to the symmetric and antisymmetric operators as $\hat{\tau}_x=-\hat{\tau}_a$ and $\hat{\tau}_y=-i\hat{\tau}_s$.

\begin{figure*}[t]
    \centering
    \includegraphics[width=0.99\linewidth]{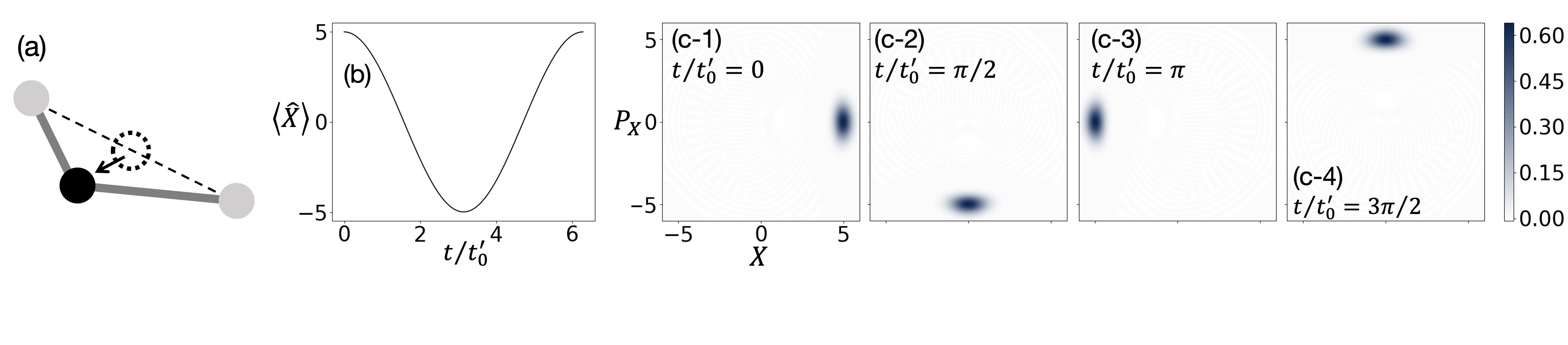}
    \vspace{-8mm}
    \caption{
    (a) Sketch of displacing the centre atom of a linear triatomic molecule. 
    Time evolution of (b) displacement $\langle\hat{X}\rangle$ and (c) the Wigner quasiprobability distribution on phase space $(X,P_X)$ for $N=50$. 
    The simulator Hamiltonian is given by $-\alpha\hat{N}_0$, and time is normalised by $t_0'=1/\alpha$.
    }
    \label{fig:simpleoscillation}
\end{figure*}

The $SU(2)$ subalgebra of mode $x$ is then generated by
\begin{subequations} \label{eq:SU2_modex}
\begin{align}
    \hat{X}_{\rm x}
    &\equiv
    \frac{
    \hat{\sigma}^{\dagger}\hat{\tau}_{x}+\hat{\tau}_{x}^{\dagger}\hat{\sigma}
    }{2}
    =
    -\frac{\hat{Q}_{zx}}{2}
    ,
    \\
    \hat{X}_{\rm y}
    &\equiv
    \frac{\hat{\sigma}^{\dagger}\hat{\tau}_{x}-\hat{\tau}_{x}^{\dagger}\hat{\sigma}}{2i}
    =
    -\frac{\hat{J}_y}{2}
    ,
    \\
    \hat{X}_{\rm z}
    &\equiv
    \frac{\hat{\sigma}^{\dagger}\hat{\sigma}-\hat{\tau}_{x}^{\dagger}\hat{\tau}_{x}}{2}
    =
    \frac{
    -\sqrt{3}\hat{Y} + \hat{D}_{xy}
    }{4}
    ,
\end{align}
\end{subequations}
while the mode $y$ subspace is spanned by
\begin{subequations}
\begin{align}
    \hat{Y}_{\rm x}
    &\equiv
    \frac{\hat{\sigma}^{\dagger}\hat{\tau}_{y}+\hat{\tau}_{y}^{\dagger}\hat{\sigma}}{2}
    =
    \frac{\hat{Q}_{yz}}{2}
    ,
    \\
    \hat{Y}_{\rm y}
    &\equiv
    \frac{\hat{\sigma}^{\dagger}\hat{\tau}_{y}-\hat{\tau}_{y}^{\dagger}\hat{\sigma}}{2i}
    =
    -\frac{\hat{J}_x}{2}
    ,
    \\
    \hat{Y}_{\rm z}
    &\equiv
    \frac{\hat{\sigma}^{\dagger}\hat{\sigma}-\hat{\tau}_{y}^{\dagger}\hat{\tau}_{y}}{2}
    =
    \frac{
    -\sqrt{3}\hat{Y} - \hat{D}_{xy}
    }{4}
    .
\end{align}
\end{subequations}
For completeness, we introduce one more subalgebra, 
\begin{subequations}
\begin{align}
    \hat{L}_{\rm x}
    &\equiv
    \frac{\hat{\tau}_+^{\dagger}\hat{\tau}_{-}+\hat{\tau}_{-}^{\dagger}\hat{\tau}_{+}}{2}
    =
    \frac{\hat{D}_{xy}}{2}
    ,
    \\
    \hat{L}_{\rm y}
    &\equiv
    \frac{\hat{\tau}_+^{\dagger}\hat{\tau}_{-}-\hat{\tau}_{-}^{\dagger}\hat{\tau}_{+}}{2i}
    =
    \frac{\hat{Q}_{xy}}{2}
    ,
    \\
    \hat{L}_{\rm z}
    &\equiv
    \frac{\hat{\tau}_+^{\dagger}\hat{\tau}_+-\hat{\tau}_{-}^{\dagger}\hat{\tau}_{-}}{2}
    =
    \frac{\hat{J}_z}{2}
    .
\end{align}
\end{subequations}
Note that the operators ${\hat{J}_x,\hat{J}_y,\hat{J}_z}$ have the following relation with
these quasi-spin-1/2 operators, $\hat{J}_x=-2\hat{Y}_{\rm y}$, $\hat{J}_y=-2\hat{X}_{\rm y}$, $\hat{J}_z=2\hat{L}_{\rm z}$. 
These operators $\hat{X}_i,\hat{Y}_i,\hat{L}_i$ are also related to each other via collective unitary transformations, for instance $\hat{U}^{\dagger}\hat{X}_{i}\hat{U}=\hat{Y}_{i}$ for $i=\{x,y,z\}$ for some unitary transformation $\hat{U}$.

\section{Bending dynamics under the linear Hamiltonian} \label{app:linear2bent}

As a complement to Sec.~\ref{sec:quench}, we also consider a linear triatomic molecule and move the centre atom (see Fig.~\ref{fig:simpleoscillation}(a)). This provides a simple illustration of how the coherent state describes the evolution of the molecule. To this end, we take the coherent state~\eqref{eq:cohernetstate_1} for $\theta=\pi/2$ as the initial state and observe the time evolution according to the Hamiltonian for a strictly linear molecule. It corresponds to $-\hat{N}_0$ according to Eq.~\eqref{eq:H2d_essen} for $\gamma=0$. We observe the Wigner quasiprobability distribution to understand how it reflects the movements of the centre atom of the molecule. We assume $-\alpha\hat{N}_0$ for the simulator Hamiltonian and normalise time with $t_0'=1/\alpha$.
Figure~\ref{fig:simpleoscillation}(b,c) shows the displacement $\langle\hat{X}\rangle$ and the Wigner quasiprobability distribution on phase space $(X,P_X)$ over time for $N=50$. 
As expected, we observe a simple harmonic oscillation of the centre atom along the $x$ coordinate.

\section{Low-depletion approximation}\label{app:lowdeplete}

A frequently used description of the spinor BEC dynamics is based on the low-depletion approximation. However, as we will see, this approximation renders the Hamiltonian independent of the parameter $\gamma$ that controls the quantum phase of the system and thus makes it impossible to observe the bending transition. The low-depletion approximation reflects the experimental limitations, particularly related to phase noise~\cite{Peise2015Interaction}, that limits the observable time evolution to the early times where the particle number $N_0$ in mode 0 is significant~\cite{Kruse2016Improvement,Hamley2012Spin,Linnemann2016Quantum}. The approximation thus is based on the assumption that $N_0\approx N\gg 1$ and one replaces $\hat{\tau}_0$, $\hat{\tau}_0^{\dagger}$ with $\sqrt{N}$~\cite{Feldmann2018Interferometric,FeldmannThesis,Pezze2019Heralded}. Under this approximation, the motion in $x$ and $y$ decouples, as the Hamiltonian reads
\begin{align} \label{eq:H_appro}
    \frac{\hat{H}_{LD}}{c/2}
    &\approx
    \hat{H}_{x}
    +
    \hat{H}_{y}
    ,
\end{align}
where constant terms are dropped and 
\begin{align}
    \hat{H}_{j}
    &=
    -2\hat{N}_{j}
    +
    \hat{\tau}_{j}^2
    +
    \hat{\tau}_{j}^{\dagger 2}
    \quad\text{for }j=x,y
    .
\end{align}
Each of the $\hat{H}_{x}, \hat{H}_y$ expresses a single-mode squeezing operator and generates entanglement. 
In the remaining quantum modes, the initial state is the vacuum state $\ket{\phi}=\ket{0}_{x}\otimes\ket{0}_{y}$. Since $[\hat{H}_{x},\hat{H}_{y}]=0$, the time-evolved state remains a product state as $e^{-i\hat{H}_{x}t}\ket{0}_x\otimes e^{-i\hat{H}_{y}t}\ket{0}_y$. 

Although the low-depletion approximation predicts entanglement generation in the mode $x,y$ subspaces, which may be simulated and explored in its own right, it drops the $\gamma$-dependency.
The phase transition of the Hamiltonian~\eqref{eq:H_spinor_2DVM} occurs at $\gamma=\gamma_c$ as mentioned in Sec.~\ref{sec:2DVM}, but the approximated Hamiltonian~\eqref{eq:H_appro} is insensitive to this transition. We therefore avoid the low-depletion approximation in our study of the phase transition.

\section{Stability of the mean-field ground state for finite $N$} \label{sec:finiteN}

We have considered dynamical changes such as bending a linear molecule and straightening a bent molecule. Here, we consider more subtle cases for complement.
First, we consider quasilinear molecules, characterised by $0<\gamma<\gamma_c$.
For $\gamma=0$, the lowest energy state is the vacuum state $\ket{0,0,0}$ for any $N$. Even for $\gamma\neq 0$, the vacuum state $\ket{0,0,0}$ is the lowest energy state in the mean-field limit as long as $\gamma\leq\gamma_c$ as mentioned in Sec.~\ref{sec:2DVM}. However, that is not the case for finite $N$ exactly. 
As shown in Fig.~\ref{fig:squeezing_QFI_time}(a), if the initial state is $\ket{0,0,0}$ and evolves according to the Hamiltonian~\eqref{eq:H_spinor_2DVM} for $\gamma=0.1<\gamma_c$, slight entanglement is generated. Nevertheless, it is much weaker than that of $\gamma=0.3>\gamma_c$.

\begin{figure}[t]
    \centering
    \includegraphics[width=0.99\linewidth]{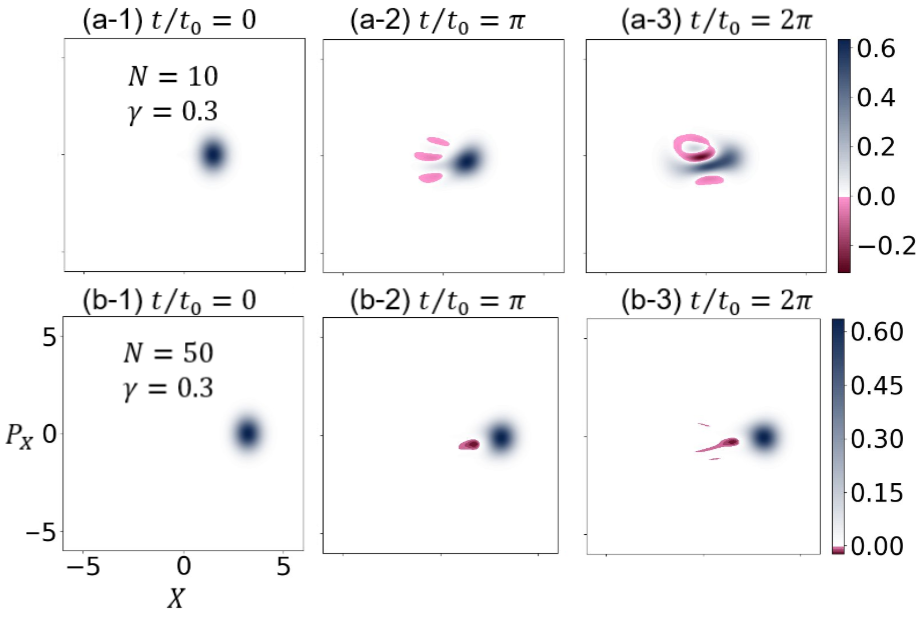}
    \caption{
    Time evolution of the Wigner quasiprobability distribution in phase space $(X,P_X)$, where the initial state is given by the coherent state~\eqref{eq:cohernetstate} for $x=r_0$, $y=0$, and $\gamma=0.3$. In (a), $N=10$ is used, and in (b) $N=50$ is used.
    }
    \label{fig:wigner_bent_N10_50}
\end{figure}

As mentioned in Sec.~\ref{sec:2DVM}, the bending degree is given by $r_0$, Eq.~\eqref{eq:r0}, in the mean-field limit when $\gamma>\gamma_c$, and the lowest energy state is given by the coherent state~\eqref{eq:cohernetstate} with $r=r_0$. 
However, this is also not true for finite $N$. 
As an example, we consider $\gamma=0.3>\gamma_c$ and the coherent state~\eqref{eq:cohernetstate} for $x=r_0$ and $y=0$ as the initial state and observe the time evolution according to the Hamiltonian~\eqref{eq:H_spinor_2DVM}. Figure~\ref{fig:wigner_bent_N10_50}(a) shows the time evolution of the Wigner quasiprobability in phase space $(X,P_X)$ for $N=10$, where a significant amount of the negativity appears. This indicates that the initial state is not the ground state in this case. Such mismatch between the finite $N$ case and the mean field limit is expected to be less visible for larger $N$. For instance, by increasing $N$ to $50$, the negativity becomes less significant (see Fig.~\ref{fig:wigner_bent_N10_50}(b)). 

\end{document}